\documentstyle[preprint,prc,eqsecnum,aps,epsf]{revtex}

\tightenlines

\newcommand{\ffat}[1]{\mbox {\boldmath $#1$}}

\begin{document}
\draft

\title{The Nd Break-Up Process in Leading Order in a Three-Dimensional Approach}

\author{ I.~Fachruddin$^{\dagger}$\thanks{ Permanent address: Jurusan Fisika, 
Universitas Indonesia, Depok 16424, Indonesia}, Ch.~Elster$^{\ddagger}$,
W.~Gl\"ockle$^{\dagger}$} 
\address{ $^{\dagger}$Institut f\"ur Theoretische Physik II, 
Ruhr-Universit\"at Bochum, D-44780 Bochum, Germany.}
\address{$^{\ddagger}$ Department of Physics and Astronomy, Ohio University, Athens, OH 34701,
and Institut f\"ur Kernphysik, Forschungszentrum J\"ulich, D-52425 J\"ulich. }

\vspace{10mm}

\date{\today}

\maketitle

\begin{abstract}
A three-dimensional approach based on momentum vectors as variables for solving the three nucleon
Faddeev equation in first order is presented. 
The nucleon-deuteron break-up amplitude is evaluated in leading order in the NN T-matrix, 
which is also generated
directly in three dimensions avoiding a summation of partial wave contributions. 
A comparison of semi-exclusive
observables in the $d(p,n)pp$ reaction calculated in this scheme with those generated by a traditional
partial wave expansion shows perfect agreement at lower energies. At about 200 MeV nucleon laboratory
energies deviations in the peak of the cross section appear, which may indicate that special care is
required in a partial wave approach for energies at and higher than 200 MeV. The role of higher order
rescattering processes beyond the leading order in the NN T-matrix is investigated with the result,
that at 200 MeV rescattering still provides important contributions 
to the cross section and certain spin
observables. The influence of a relativistic treatment of the kinematics is investigated.
It is found that relativistic effects become important at projectile energies higher than
200~MeV.
\end{abstract}

\vspace{10mm}


\pagebreak
 
\narrowtext


\section{Introduction}

During the last two decades calculations of nucleon-deuteron (Nd) scattering based on momentum space 
Faddeev equations \cite{faddeev} experienced enormous improvement and refinement. It is fair 
to state that below about 200 MeV projectile energy the momentum space Faddeev equations for 
three-nucleon (3N) scattering now can be solved with very high accuracy for the most modern two and 
three nucleon forces. A summary of these achievements is given in Ref.~\cite{1}. The approach
to 3N scattering described in  Ref.~\cite{1} is based on using angular momentum eigenstates for the
two- and three-body system. 
For low projectile energies this procedure is certainly physically justified due to arguments related to 
the centrifugal barrier. However, to probe the strong interaction at shorter distances one has to go 
to higher projectile energies, where the algebraic and algorithmic work carried out in traditional 
partial wave (PW) decomposition can be 
quite involved. A more crucial hurdle is posed by the fact that in three-nucleon scattering (3N)
calculations for 
projectile energies of a few hundred MeV  the number of partial waves needed to achieve 
numerical convergence proliferates, and limitations with respect to computational feasibility and 
accuracy are being reached. 
It appears therefore natural to abandon PW representations completely and work directly with vector 
variables, if one wants to calculate 3N scattering at higher energies.  
As an aside, the use of vector variables is  common practice in bound state calculations 
based on variational \cite{variation} and Green's Function Monte Carlo (GFMC) methods \cite{12}, 
which are carried out in coordinate space. 

Momentum space calculations within the Faddeev scheme which did not employ a PW decomposition were
first carried out for a system of three bosons~\cite{trit3d,scatter1}. 
Here the momentum space Faddeev equations were solved for the bound as well as the scattering state. 
In this work
we want to employ realistic nucleon-nucleon (NN) interactions in a 3N scattering calculation.
This means we have to incorporate spin degrees of freedom into the formulation of the Faddeev equations. 
Since the input to any Faddeev calculation is the solution of the Lippmann-Schwinger (LS) equation 
for the two-nucleon T-matrix, we start from the formulation of NN scattering developed in Ref.~\cite{nn3d}. 
There we chose an approach based on the total helicity of the NN system as spin variable. 
From our point of view this is the preferred starting point to later progress to the 3N system. 

In this work we consider the first term of the multiple scattering series built up by the Faddeev
equations, rather than solve the full  Faddeev equations for three nucleons, and 
concentrate on semi-exclusive break-up observables. 
Of particular interest are the polarization-transfer coefficients in the (p,n) charge exchange reaction on 
the deuteron, which recently have been measured at IUCF with a projectile energy of 197 MeV \cite{ndx197} 
and at RCNP with a projectile energy of 346 MeV\cite{ndx346}. Since these measurements are carried out 
at 'intermediate energies', one can speculate that it may be sufficient to consider only the first 
order term in the multiple scattering series. Furthermore, since the projectile energies are high, 
we will consider relativistic effects as far as the kinematics is concerned.

In Section II we formulate the Nd break-up process in a three-dimensional (3D), nonrelativistic 
Faddeev scheme. We derive the leading term of the full Nd break-up amplitude, where NN T-matrix 
elements are given in the momentum-helicity basis defined in Ref.~\cite{nn3d}. In Section III 
we introduce relativistic kinematics into this formulation. We will not consider a  boost of the 
NN T-matrix \cite{kamada} nor Wigner's rotations \cite{polyzou} of the spin. 
The observables for the (p,n) charge exchange reaction are introduced in Section IV.
In Section V we present and discuss our calculations for the (p,n) charge exchange reaction in the
proton-deuteron (pd) break-up process. Here only the outgoing neutron is detected after the
break-up. We present calculations of  spin averaged differential cross sections, neutron
polarizations, proton analyzing powers, and polarization-transfer coefficients at different energies.
We also compare our calculations with traditional PW calculations.
Finally, we summarize in Section VI.


\section{The Nonrelativistic Nd Break-Up Amplitude}

In the Faddeev scheme the operator $U_0^{full}$ for the Nd break-up process is given as \cite{1} 
\begin{equation}
U_0^{full} = (1 + P)T_F .
\label{eq:2.1}
\end{equation}
Here $T_F$ is the Faddeev operator obeying the Faddeev equation \cite{faddeev} for the break-up process of 
three identical particles, 
\begin{equation}
\label{nd6}
T_F = TP + TG_0PT_F .
\end{equation}
The operator $T$ stands for the NN t-matrix, and $P$ is a permutation operator defined as
\begin{equation}\label{nd1}
P \equiv P_{12}P_{23} + P_{13}P_{23}.
\end{equation}
The free three-nucleon (3N) propagator is given by $G_0$. The matrix elements of the
break-up amplitude $U_0^{full}({\bf p},{\bf q})$ of Eq.~(\ref{eq:2.1}) are defined as
\begin{eqnarray}
U_0^{full}({\bf p},{\bf q}) & \equiv & \left\langle {\bf p}{\bf q}m_{s1}m_{s2}m_{s3}\tau_1\tau_2\tau_3 \biggl| U_0^{full} \biggr| {\bf q}_0m_{s1}^0\tau_1^0 \Psi _d^{M_d}\right\rangle \nonumber\\
& = & \left\langle {\bf p}{\bf q}m_{s1}m_{s2}m_{s3}\tau_1\tau_2\tau_3 \biggl| (1+P)T_F \biggr| 
{\bf q}_0m_{s1}^0\tau_1^0 \Psi _d^{M_d}\right\rangle ,
\label{nd5} 
\end{eqnarray}
where 
\begin{equation}
|{\bf p}{\bf q}m_{s1}m_{s2}m_{s3}\tau_1\tau_2\tau_3 \rangle \equiv |{\bf q}m_{s1}\tau_1 \rangle 
|{\bf p}m_{s2}m_{s3}\tau_2\tau_3 \rangle  
\label{eq:2.5}
\end{equation}
is the final, not-antisymmetrized free 3N state. 
The quantities $m_{si},\tau_i$ ($i = 1,2,3$) are the spins and isospins of the three nucleons.
The initial state, in which only the deuteron state $\left|\Psi _d^{M_d}\right\rangle $ is
antisymmetrized, is given by
\begin{equation}
\left| {\bf q}_0m_{s1}^0\tau_1^0 \Psi _d^{M_d}\right\rangle \equiv \left| {\bf q}_0m_{s1}^0\tau_1^0 
\right\rangle \left|\Psi _d^{M_d}\right\rangle 
\label{eq:2.6}
\end{equation}
The index  $M_d$ indicates the projection of total angular momentum of the deuteron
 along an arbitrary z-axis, 
$m_{s1}^0,\tau_1^0$ are the spin and isospin of nucleon `1' acting as the projectile. 
Without loss of generality nucleon `1' is singled out as projectile, while the other two nucleons, 
`2' and `3' form the two-nucleon (2N) subsystem, i.e. the deuteron in the initial state.
Jacobi momenta ${\bf p}$ and ${\bf q}$ are used to describe the 3N kinematics in the final state,
\begin{eqnarray}
{\bf p} & = & \frac{1}{2}\left( {\bf k}_{2}-{\bf k}_{3}\right)\\
{\bf q} & = & \frac{2}{3}\left[ {\bf k}_{1}-\frac{1}{2}\left( {\bf k}_{2}+{\bf k}_{3}\right) \right] = 
{\bf k}_{1} - \frac{1}{3}{\bf k}_{lab} \label{nda1}.
\end{eqnarray}
Here the ${\bf k}_i$'s ($i = 1,2,3$) represent the laboratory momenta of the three nucleons.
Defining ${\bf k}_{lab}$ as the laboratory momentum of the projectile and 
applying momentum conservation 
\begin{equation}
{\bf k}_{lab} = {\bf k}_{1}+{\bf k}_{2}+{\bf k}_{3} ,
\end{equation}
leads to  Eq.~(\ref{nda1}).
In the initial state ${\bf q}_0$ is the relative momentum of the projectile to the target deuteron and is
related to ${\bf k}_{lab}$ as
\begin{eqnarray}
{\bf q}_0 & = & \frac{2}{3}{\bf k}_{lab} .
\end{eqnarray}
For clarity of description we will in the following denote the break-up amplitude as 
$U_0^{full}({\bf p},{\bf q})$ and suppress all other quantum numbers. 

In this work we only want to consider the leading term 
of the full break-up operator $U_0^{full}$. This means we only consider the leading term
in the  Faddeev operator $T_F$ of Eq.~(\ref{nd6}) and define the break-up operator $U_0$ in first 
order in the NN T-matrix as
\begin{equation}
U_0 = (1 + P)TP .
\label{eq:2.11}
\end{equation}
The matrix elements of $U_0({\bf p},{\bf q})$ with respect to the final and initial states from
Eqs.~(\ref{eq:2.5}) and (\ref{eq:2.6}) are then given as
\begin{eqnarray}
U_0({\bf p},{\bf q}) & \equiv & \left\langle {\bf p}{\bf q}m_{s1}m_{s2}m_{s3}\tau_1\tau_2\tau_3 
\biggl| U_0 \biggr| {\bf q}_0m_{s1}^0\tau_1^0 \Psi _d^{M_d}\right\rangle \nonumber\\ 
& = & \left\langle {\bf p}{\bf q}m_{s1}m_{s2}m_{s3}\tau_1\tau_2\tau_3 \biggl| (1 + P)TP 
\biggr| {\bf q}_0m_{s1}^0\tau_1^0 \Psi _d^{M_d}\right\rangle 
\label{nd7}.
\end{eqnarray}
From now one we mean by the Nd break-up amplitude (or break-up amplitude)
 the matrix element $U_0({\bf p},{\bf q})$ given in Eq.~(\ref{nd7}) 
and by the full Nd break-up amplitude (or full break-up amplitude) the matrix element
$U_0^{full}({\bf p},{\bf q})$ given in Eq.~(\ref{nd5}). 

The break-up amplitude $U_0({\bf p},{\bf q})$ from Eq.~(\ref{nd7}) is composed out of three terms,  
\begin{equation}
U_{0}({\bf p},{\bf q}) = U_{0}^{(1)}({\bf p},{\bf q})+U_{0}^{(2)}({\bf p},{\bf q})+U_{0}^{(3)}({\bf p},{\bf q}) ,
\end{equation}
with
\begin{eqnarray}
U_{0}^{(1)}({\bf p},{\bf q}) & \equiv  & { \atop } _1 \left\langle {\bf p}{\bf q}m_{s1}m_{s2}m_{s3}\tau_1\tau_2\tau_3 \biggl| TP \biggr| {\bf q}_0m_{s1}^0\tau_1^0 \Psi _d^{M_d}\right\rangle \label{nd11}\\
U_{0}^{(2)}({\bf p},{\bf q}) & \equiv  & { \atop } _1 \left\langle {\bf p}{\bf q}m_{s1}m_{s2}m_{s3}\tau_1\tau_2\tau_3 \biggl| P_{12}P_{23}TP \biggr| {\bf q}_0m_{s1}^0\tau_1^0 \Psi _d^{M_d}\right\rangle \label{nd8}\\
U_{0}^{(3)}({\bf p},{\bf q}) & \equiv  & { \atop } _1 \left\langle {\bf p}{\bf q}m_{s1}m_{s2}m_{s3}\tau_1\tau_2\tau_3 \biggl| P_{13}P_{23}TP \biggr| {\bf q}_0m_{s1}^0\tau_1^0 \Psi _d^{M_d}\right\rangle .
\end{eqnarray}
Here the final free 3N state are labeled  `1', meaning that nucleons `2' and `3' form the 2N-subsystem. 
The spin and isospin quantum numbers must be read in the order 123. 
Applying the  permutation operator $P_{12}P_{23}$ to the final state of  $U_{0}^{(2)}({\bf p},{\bf q})$
given in Eq.~(\ref{nd8}) leads to
\begin{eqnarray}
U_{0}^{(2)}({\bf p},{\bf q}) & = & { \atop } _1 \left\langle P_{23}P_{12}{\bf p}{\bf q}m_{s1}m_{s2}m_{s3}
\tau_1\tau_2\tau_3 \biggl| TP \biggr| {\bf q}_0m_{s1}^0\tau_1^0 \Psi _d^{M_d}\right\rangle \nonumber\\
& = & { \atop } _3 \left\langle {\bf p}{\bf q}m_{s1}m_{s2}m_{s3}\tau_1\tau_2\tau_3 \biggl| TP 
\biggr| {\bf q}_0m_{s1}^0\tau_1^0 \Psi _d^{M_d}\right\rangle.
\label{eq:2.17}
\end{eqnarray}
This is now the final state where  nucleons `1' and `2' form the 2N subsystem. Accordingly,
 the spin and isospin quantum numbers associated with the three nucleons read in the order 312. 
In order to have the same final states as $U_{0}^{(1)}({\bf p},{\bf q})$ in Eq.~(\ref{nd11}), we
need to transform the final state such that nucleons `2' and `3' form the 2N subsystem. 
This transformation is achieved by the following relation among the Jacobi momenta \cite{book}
\begin{equation}\label{nd3}
{\bf p}_1 = -\frac{1}{2}{\bf p}_3 - \frac{3}{4}{\bf q}_3 
\qquad \qquad {\bf q}_1 = {\bf p}_3 - \frac{1}{2}{\bf q}_3 ,
\end{equation}
where the labels 1 and 3 indicate the nucleon being singled out. This leads to
\begin{equation}
U_{0}^{(2)}({\bf p},{\bf q}) = { \atop } _1 \left\langle \left(-\frac{1}{2}{\bf p}-
\frac{3}{4}{\bf q}\right)\left({\bf p}-\frac{1}{2}{\bf q}\right)m_{s2}m_{s3}m_{s1}\tau_2\tau_3\tau_1 
\biggl| TP \biggr| {\bf q}_0m_{s1}^0\tau_1^0 \Psi _d^{M_d}\right\rangle .
\label{eq:2.19}
\end{equation}
Using another relation between Jacobi momenta \cite{book} 
\begin{equation}\label{nd4}
{\bf p}_1 = -\frac{1}{2}{\bf p}_2 + \frac{3}{4}{\bf q}_2 \qquad \qquad {\bf q}_1 = -{\bf p}_2 - 
\frac{1}{2}{\bf q}_2 ,
\end{equation}
we can obtain  $U_{0}^{(3)}({\bf p},{\bf q})$ in a similar fashion as 
\begin{equation}
U_{0}^{(3)}({\bf p},{\bf q}) = { \atop } _1 \left\langle \left(-\frac{1}{2}{\bf p}+\frac{3}{4}{\bf q}\right)
\left(-{\bf p}-\frac{1}{2}{\bf q}\right)m_{s3}m_{s1}m_{s2}\tau_3\tau_1\tau_2 
\biggl| TP \biggr| {\bf q}_0m_{s1}^0\tau_1^0 \Psi _d^{M_d}\right\rangle .
\label{eq:2.21}
\end{equation}
Since $U_{0}^{(2)}({\bf p},{\bf q})$ and $U_{0}^{(3)}({\bf p},{\bf q})$ differ from $U_{0}^{(1)}({\bf p},{\bf q})$ 
only in their variables,
it is sufficient to work out an expression only for one of them, which we choose to be 
$U_{0}^{(1)}({\bf p},{\bf q})$. For calculating subsequently  $ U_{0}^{(2)}({\bf p},{\bf q})$ and 
$ U_{0}^{(3)}({\bf p},{\bf q})$ one only has to perform the following replacements,
\begin{eqnarray}
\mbox{for }U_{0}^{(2)}({\bf p},{\bf q}) & : & \{\tau ,m\}_{\{1,2,3\}} \rightarrow \{\tau ,m\}_{\{2,3,1\}} \qquad {\bf p} \rightarrow -\frac{1}{2}{\bf p}-\frac{3}{4}{\bf q} \qquad {\bf q} \rightarrow {\bf p}-\frac{1}{2}{\bf q} \\
\mbox{for }U_{0}^{(3)}({\bf p},{\bf q}) & : & \{\tau ,m\}_{\{1,2,3\}} \rightarrow \{\tau ,m\}_{\{3,1,2\}} \qquad {\bf p} \rightarrow -\frac{1}{2}{\bf p}+\frac{3}{4}{\bf q} \qquad {\bf q} \rightarrow -{\bf p}-\frac{1}{2}{\bf q} . 
\end{eqnarray}

For calculating $U_{0}^{(1)}({\bf p},{\bf q})$ we start
by inserting the following completeness relation for the free 3N system
\begin{equation}
\sum_{{m_{s1}m_{s2}m_{s3} \atop \tau_1\tau_2\tau_3}} \int d{\bf p} \int d{\bf q} \, |{\bf p}{\bf q}m_{s1}m_{s2}m_{s3}\tau_1\tau_2\tau_3 \rangle \langle{\bf p}{\bf q}m_{s1}m_{s2}m_{s3}\tau_1\tau_2\tau_3 | = 1
\end{equation}
twice into Eq.~(\ref{nd11}), which leads to
\begin{eqnarray}
U_{0}^{(1)}({\bf p},{\bf q}) & = & \sum _{{m_{s1}'m_{s2}'m_{s3}' \atop \tau _{1}'\tau _{2}'\tau _{3}'}} \int d{\bf p}' \int d{\bf q}'\left\langle {\bf p}{\bf q}m_{s1}m_{s2}m_{s3}\tau _{1}\tau _{2}\tau _{3}\right| T \left| {\bf p}'{\bf q}'m_{s1}'m_{s2}'m_{s3}'\tau _{1}'\tau _{2}'\tau _{3}'\right\rangle \nonumber\\
 &  & \times \sum _{{m_{s1}''m_{s2}''m_{s3}'' \atop \tau _{1}''\tau _{2}''\tau _{3}''}} \int d{\bf p}'' \int d{\bf q}''\left\langle {\bf p}'{\bf q}'m_{s1}'m_{s2}'m_{s3}'\tau _{1}'\tau _{2}'\tau _{3}'\right| P\left| {\bf p}''{\bf q}''m_{s1}''m_{s2}''m_{s3}''\tau _{1}''\tau _{2}''\tau _{3}''\right\rangle \nonumber\\
 &  & \times \left\langle {\bf p}''{\bf q}''m_{s1}''m_{s2}''m_{s3}''\tau _{1}''\tau _{2}''\tau _{3}'' \right. \left| {\bf q}_0m_{s1}^0\tau_1^0 \Psi _d^{M_d}\right\rangle \nonumber\\
& = & \sum _{m_{s2}'m_{s3}'\tau _{2}'\tau _{3}'} \int d{\bf p}' \left\langle \left.{\bf p}m_{s2}m_{s3}\tau _{2}\tau _{3}\right. \left| T (E_p) \right| \left. {\bf p}'m_{s2}'m_{s3}'\tau _{2}'\tau _{3}'\right. \right\rangle \nonumber\\
 &  & \times \sum _{m_{s2}''m_{s3}''\tau _{2}''\tau _{3}''}\int d{\bf p}'' \left\langle {\bf p}'m_{s2}'m_{s3}'\tau _{2}'\tau _{3}'\right| \left\langle {\bf q}m_{s1}\tau _{1}\left| P \right| {\bf q}_0m_{s1}^0\tau _{1}^0\right\rangle \left| {\bf p}''m_{s2}''m_{s3}''\tau _{2}''\tau _{3}''\right\rangle \nonumber\\
 &  & \times \left\langle {\bf p}''m_{s2}''m_{s3}''\tau _{2}''\tau _{3}''\right. \left| \Psi _d^{M_d}\right\rangle \label{nd13}.
\end{eqnarray}

In arriving at the last equality we used the fact that $T$ acts only in the two-particle
subsystem together with  momentum space properties of the initial state.

The NN T-matrix is calculated at a center of mass (c.m.) energy $E_p$ of the 23-subsystem 
\begin{equation}
E_p \equiv \frac{p^2}{m} = \frac{3}{4m}(q_0^2-q^{2})+E_d \label{nd55},
\end{equation}
which does not necessarily corresponds to the intermediate relative momenta ${\bf p}'$. 
The deuteron binding energy is represented by $E_d$, and $m$ stand for the nucleon mass.

Using the relations for Jacobi momenta given in Eqs.~(\ref{nd3}) and (\ref{nd4}) the evaluation 
of the permutations in Eq.~(\ref{nd13}) leads to 
\begin{eqnarray}
\lefteqn {\left\langle {\bf p}'m_{s2}'m_{s3}'\tau _{2}'\tau _{3}'\right| \left\langle {\bf q}m_{s1}\tau _{1}\left| P \right| {\bf q}_0m_{s1}^0\tau _{1}^0\right\rangle \left| {\bf p}''m_{s2}''m_{s3}''\tau _{2}''\tau _{3}''\right\rangle } &  & \nonumber\\
 &  & \qquad =\, {_1\left\langle {\bf p}'\right|} \left\langle {\bf q}\right. \left| {\bf q}_0\right\rangle \left| {\bf p}''\right\rangle _2 {{\atop}_1 \left\langle m_{s1}m_{s2}'m_{s3}'\tau _{1}\tau _{2}'\tau _{3}'\left. {\atop}\right| m_{s1}^0m_{s2}''m_{s3}''\tau _{1}^0\tau _{2}''\tau _{3}''\right\rangle _2} \nonumber\\
 &  & \qquad \quad + {_1\left\langle {\bf p}'\right|} \left\langle {\bf q}\right. \left| {\bf q}_0\right\rangle \left| {\bf p}''\right\rangle _3 {{\atop}_1 \left\langle m_{s1}m_{s2}'m_{s3}'\tau _{1}\tau _{2}'\tau _{3}'\left. {\atop}\right| m_{s1}^0m_{s2}''m_{s3}''\tau _{1}^0\tau _{2}''\tau _{3}''\right\rangle _3} \nonumber\\
 &  & \qquad =\, {\atop}_{1}\left\langle {\bf p}'\left| \left\langle {\bf q}\biggl. \biggr| -{\bf p}''-\frac{1}{2}{\bf q}_0\right\rangle \right| -\frac{1}{2}{\bf p}''+\frac{3}{4}{\bf q}_0\right\rangle _{1} {{\atop}_{1}\left\langle m_{s1}m_{s2}'m_{s3}'\tau _{1}\tau _{2}'\tau _{3}'\left. {\atop}\right| m_{s3}''m_{s1}^0m_{s2}''\tau _{3}''\tau _{1}^0\tau _{2}''\right\rangle _1} \nonumber\\
 &  & \qquad \quad + {\atop}_{1}\left\langle {\bf p}'\left| \left\langle {\bf q}\biggl. \biggr| {\bf p}''-\frac{1}{2}{\bf q}_0\right\rangle \right| -\frac{1}{2}{\bf p}''-\frac{3}{4}{\bf q}_0\right\rangle _{1} {{\atop}_{1}\left\langle m_{s1}m_{s2}'m_{s3}'\tau _{1}\tau _{2}'\tau _{3}'\left. {\atop}\right| m_{s2}''m_{s3}''m_{s1}^0\tau _{2}''\tau _{3}''\tau _{1}^0\right\rangle _1} \nonumber\\
 & & \qquad =\,\delta \left( {\bf p}'-{\ffat \pi }\right) \delta \left( {\bf p}''-{\ffat \pi }'\right) \delta _{m_{s1}m_{s3}''}\delta _{m_{s2}'m_{s1}^0}\delta _{m_{s3}'m_{s2}''}\delta _{\tau _{1}\tau _{3}''}\delta _{\tau _{2}'\tau _{1}^0}\delta _{\tau _{3}'\tau _{2}''} \nonumber\\
 &  & \qquad \quad + \delta \left( {\bf p}'+{\ffat \pi }\right) \delta \left( {\bf p}''+{\ffat \pi }'\right) \delta _{m_{s1}m_{s2}''}\delta _{m_{s2}'m_{s3}''}\delta _{m_{s3}'m_{s1}^0}\delta _{\tau _{1}\tau _{2}''}\delta _{\tau _{2}'\tau _{3}''}\delta _{\tau _{3}'\tau _{1}^0} \label{nd12}, \quad 
\end{eqnarray}
where
\begin{eqnarray}
{\ffat \pi }\equiv \frac{1}{2}{\bf q}+{\bf q}_0 & \qquad \qquad & 
{\ffat \pi }'\equiv -{\bf q}-\frac{1}{2}{\bf q}_0 .
\end{eqnarray}
As an aside, the variables are arranged such that each delta function only contains one integration
variable.  Inserting Eq.~(\ref{nd12})
into $U_0^{(1)}({\bf p},{\bf q})$ in Eq.~(\ref{nd13}) leads to
 \begin{eqnarray}
U_{0}^{(1)}({\bf p},{\bf q}) & = & \sum _{m_{s3}'\tau _{3}'} \left\langle {\bf p}m_{s2}m_{s3}\tau _{2}\tau _{3} \left| T (E_p) \right| {\ffat \pi }m_{s1}^0m_{s3}'\tau _{1}^0\tau _{3}' \right\rangle \left\langle {\ffat \pi }'m_{s3}'m_{s1}\tau _{3}'\tau _{1}\left. \right| \Psi _d^{M_d}\right\rangle \nonumber\\
 &  & + \sum _{m_{s2}'\tau _{2}'} \left\langle {\bf p}m_{s2}m_{s3}\tau _{2}\tau _{3} \left| T (E_p) \right| -{\ffat \pi }m_{s2}'m_{s1}^0\tau _{2}'\tau _{1}^0 \right\rangle \left\langle -{\ffat \pi }'m_{s1}m_{s2}'\tau _{1}\tau _{2}'\left. \right| \Psi _d^{M_d}\right\rangle \nonumber\\
& = & \sum _{m_{s}'\tau '} \biggl\{ \left\langle {\bf p}m_{s2}m_{s3}\tau _{2}\tau _{3} \left| T (E_p) \right| {\ffat \pi }m_{s1}^0m_{s}'\tau _{1}^0\tau ' \right\rangle \left\langle {\ffat \pi }'m_{s}'m_{s1}\tau '\tau _{1}\left. \right| \Psi _d^{M_d}\right\rangle \nonumber\\
 &  & + \left\langle {\bf p}m_{s2}m_{s3}\tau _{2}\tau _{3} \left| T (E_p) P_{23}\right| {\ffat \pi }m_{s1}^0m_{s}'\tau _{1}^0\tau ' \right\rangle \left\langle {\ffat \pi }'m_{s}'m_{s1}\tau '\tau _{1}\left|P_{23}^{-1} \right| \Psi _d^{M_d}\right\rangle \biggr\} \nonumber\\
& = & \sum _{m_{s}'\tau '} \left\langle {\bf p}m_{s2}m_{s3}\tau _{2}\tau _{3} \left| T (E_p) (1-P_{23})\right| {\ffat \pi }m_{s1}^0m_{s}'\tau _{1}^0\tau ' \right\rangle \left\langle {\ffat \pi }'m_{s}'m_{s1}\tau '\tau _{1}\left. \right| \Psi _d^{M_d}\right\rangle \nonumber\\ 
& = & \sum _{m_{s}'\tau '} {\atop}_a\left\langle {\bf p}m_{s2}m_{s3}\tau _{2}\tau _{3} \left| T (E_p) \right| {\ffat \pi }m_{s1}^0m_{s}'\tau _{1}^0\tau ' \right\rangle _a \left\langle {\ffat \pi }'m_{s}'m_{s1}\tau '\tau _{1}\left. \right| \Psi _d^{M_d}\right\rangle \label{nd14} .
\end{eqnarray}
In arriving at 
the last equality of Eq.~(\ref{nd14}) we made use of the antisymmetry of the deuteron state,
 $\left| \Psi _d^{M_d}\right\rangle $ and defined ${\atop}_a\left\langle {\bf p}m_{s2}m_{s3}\tau _{2}
\tau _{3} \left| T (E_p) \right| {\ffat \pi }m_{s1}^0m_{s}'\tau _{1}^0\tau ' \right\rangle _a $ as 
\begin{equation}\label{nd9}
{\atop}_a\left\langle {\bf p}m_{s2}m_{s3}\tau _{2}\tau _{3} \left| T (E_p) \right| {\ffat \pi }m_{s1}^0m_{s}'\tau _{1}^0\tau ' \right\rangle _a \equiv \left\langle {\bf p}m_{s2}m_{s3}\tau _{2}\tau _{3} \left| T (E_p) (1-P_{23})\right| {\ffat \pi }m_{s1}^0m_{s}'\tau _{1}^0\tau '\right\rangle .
\end{equation}
We denote the matrix element ${\atop}_a\left\langle {\bf p}m_{s2}m_{s3}\tau _{2}\tau _{3} \left| T (E_p) \right| 
{\ffat \pi }m_{s1}^0m_{s}'\tau _{1}^0\tau ' \right\rangle _a $ as physical representation of the NN T-matrix, 
physical meaning that the NN basis states $\left| {\bf p}m_{s2}m_{s3}\tau _{2}\tau _{3} \right\rangle _a $
contain the individual spins and isospins of the nucleons.

Since the deuteron contains only two definite angular momentum states, it is reasonable to 
apply the standard partial wave expansion  
\begin{eqnarray}
\left| \Psi _d^{M_d}\right\rangle & = & \sum _{lsjmt}\int dp'\, p'^{2} \left| p'(ls)jm;t\left\rangle 
\right\langle p'(ls)jm;t\left. \right| \Psi _d^{M_d}\right\rangle \nonumber\\
& = & \sum _{l}\int dp'\, p'^{2} \left| p'(l1)1M_{d};0\right\rangle \psi _{l}(p') .
\end{eqnarray}
Here $\left| p'(ls)jm;t\right\rangle $ is the standard partial wave basis, and 
$\psi _{l}(p')$ represent the standard $s-$ and $d$-waves of the deuteron. 
The projection $\left\langle {\ffat \pi }'m_{s}'m_{s1}\tau '\tau _{1}\left. \right| \Psi _d^{M_d}\right\rangle$
on the deuteron state in Eq.~(\ref{nd14}) is then explicitly worked out as
\begin{eqnarray}
\left\langle {\ffat \pi }'m_{s}'m_{s1}\tau '\tau _{1}\left. \right| \Psi _d^{M_d}\right\rangle 
& = & \sum _{l}\int dp'\, p'^{2}\left\langle {\ffat \pi }'m_{s}'m_{s1}\tau '\tau _{1}\right. \left| p'(l1)1M_{d};0\right\rangle \psi _{l}(p')\nonumber\\
& = & \left\langle \tau '\tau _{1}\right. \left|0\right\rangle \sum _{l \mu }C(l11;\mu ,M_{d}-\mu )  \nonumber\\
&  & \times \int dp'\, p'^{2} \left\langle {\ffat \pi }'\right. \left| p'l\mu \right\rangle 
\left\langle m_{s}'m_{s1}\right. \left| 1,M_{d}-\mu \right\rangle \psi _{l}(p') \nonumber\\
& = & C\left( \frac{1}{2}\frac{1}{2}0;\tau '\tau _{1}\right) \sum _{l \mu }C(l11;\mu ,M_{d}-\mu )\nonumber\\
&  & \times Y_{l\mu }(\hat{{\ffat \pi }}')C\left( \frac{1}{2}\frac{1}{2}1;m_{s}'m_{s1}\right) \delta _{m_{s}'+m_{s1},M_{d}-\mu } \psi _{l}(\pi ') \nonumber\\
& = & C\left( \frac{1}{2}\frac{1}{2}0;\tau '\tau _{1}\right) C\left( \frac{1}{2}\frac{1}{2}1;m_{s}'m_{s1}\right) 
\nonumber\\
&  & \times \sum _{l} C\left( l11;M_{d}-m_{s}'-m_{s1},m_{s}'+m_{s1}\right) \nonumber\\
&  & \times Y_{l,M_{d}-m_{s}'-m_{s1}}(\hat{{\ffat \pi }}')\psi _{l}(\pi ') . \label{nd2}
\end{eqnarray}
Finally, inserting Eq.~(\ref{nd2}) into Eq.~(\ref{nd14}) we obtain the first part of the break-up 
amplitude $U_0^{(1)}({\bf p},{\bf q})$ as
\begin{eqnarray}
U_{0}^{(1)}({\bf p},{\bf q}) & = & \sum _{m_{s}'\tau '} {\atop}_a\left\langle {\bf p}m_{s2}m_{s3}\tau _{2}\tau _{3} \left| T (E_p) \right| {\ffat \pi }m_{s1}^0m_{s}'\tau _{1}^0\tau ' \right\rangle _a C\left( \frac{1}{2}\frac{1}{2}0;\tau '\tau _{1}\right) C\left( \frac{1}{2}\frac{1}{2}1;m_{s}'m_{s1}\right) \nonumber\\
 &  & \times \sum _{l}C\left( l11;M_{d}-m_{s}'-m_{s1},m_{s}'+m_{s1}\right) Y_{l,M_{d}-m_{s}'-m_{s1}}(\hat{{\ffat \pi }}')\psi _{l}(\pi ') \nonumber\\
& = & \frac{(-)^{\frac{1}{2}+\tau _{1}}}{\sqrt{2}}\sum _{m_{s}'} C\left( \frac{1}{2}\frac{1}{2}1;m_{s}'m_{s1}\right) {\atop}_a\left\langle {\bf p}m_{s2}m_{s3}\tau _{2}\tau _{3} \left| T (E_p) \right| {\ffat \pi }m_{s1}^0m_{s}'\tau _{1}^0\tau ' \right\rangle _a \nonumber\\
 &  & \times \sum _{l}C\left( l11;M_{d}-m_{s}'-m_{s1},m_{s}'+m_{s1}\right) Y_{l,M_{d}-m_{s}'-m_{s1}}(\hat{{\ffat \pi }}')\psi _{l}(\pi ') . \label{nd0}
\end{eqnarray}

Eq.~(\ref{nd0}) may serve as a starting point for further expressions for 
$U_{0}^{(1)}({\bf p},{\bf q})$ to be used in the explicit calculations. It shows how 
$U_0^{(1)}({\bf p},{\bf q})$ depends on the PW projected components of the deuteron and on the NN T-matrix
in a physical representation. 
In our calculation of the Nd break-up process we employ the NN t-matrix in the momentum-helicity basis 
$\left| {\bf p};\hat{{\bf p}}S\Lambda ;t\right\rangle ^{\pi a}$ \cite{nn3d}, where 
$S,\, t,\, \Lambda $ are the total spin of the 2N system, the total isospin and the helicity. The label 
$\pi a$ means that the basis state has a definite parity $\eta _{\pi }$ and is antisymmetrized. 
The connection of the T-matrix elements 
${\atop}_a\left\langle {\bf p}m_{s2}m_{s3}\tau _{2}\tau _{3} \left| T (E_p) \right| 
{\ffat \pi }m_{s1}^0m_{s}'\tau _{1}^0\tau ' \right\rangle _a $ to those in the momentum-helicity basis, 
namely $T_{\Lambda \Lambda '}^{\pi St}({\bf p},{\ffat \pi };E_p)$, is  given in Ref.~\cite{nn3d}. 
Here we want to be more general by letting the nucleon types 
$\tau _{2},\, \tau _{3},\, \tau _{1}^0,\, \tau _{1}$ being arbitrary
but employing  Kronecker symbols to ensure charge conservation, 
\begin{eqnarray}
\lefteqn { _{a}\langle \tau _{2}\tau _{3}m_{s2}m_{s3}{\bf p}| T(E_p)| \tau _{1}^0,-\tau _{1}m_{s1}^0m_{s}' {\ffat \pi }\rangle _{a}} &  & \nonumber\\
 &  & \qquad \qquad =\, \frac{1}{4}\delta _{\tau _{2}+\tau _{3},\tau _{1}^0-\tau _{1}}e^{-i(\Lambda _{0}\phi _p -\Lambda _{0}'\phi _{\pi })}\sum _{S\pi t}\left( 1-\eta _{\pi }(-)^{S+t}\right) C\left( \frac{1}{2}\frac{1}{2}t;\tau _{2}\tau _{3}\right) C\left( \frac{1}{2}\frac{1}{2}t;\tau _{1}^0,-\tau _{1}\right) \nonumber\\
 &  & \qquad \qquad \quad \times C\left( \frac{1}{2}\frac{1}{2}S;m_{s2}m_{s3}\Lambda _{0}\right) C\left( \frac{1}{2}\frac{1}{2}S;m_{s1}^0m_{s}'\Lambda _{0}'\right) \nonumber\\
 &  & \qquad \qquad \quad \times \sum _{\Lambda \Lambda '}d^{S}_{\Lambda _{0}\Lambda }(\theta _p)d^{S}_{\Lambda _{0}'\Lambda '}(\theta _{\pi })T_{\Lambda \Lambda '}^{\pi St}({\bf p},{\ffat \pi };E_p) . \label{nd10}
\end{eqnarray}
In the above expression  $d^{S}_{\Lambda '\Lambda }(\theta )$ is a rotation matrix \cite{rose}. Using Eq.~(\ref{nd10}) we obtain $U_0^{(1)}({\bf p},{\bf q})$ in terms of the NN T-matrix elements $T_{\Lambda \Lambda '}^{\pi St}({\bf p},{\ffat \pi };E_p)$ in the momentum-helicity basis as
\begin{eqnarray}
U_{0}^{(1)}({\bf p},{\bf q}) & = & \frac{(-)^{\frac{1}{2}+\tau _{1}}}{4\sqrt{2}}\delta _{\tau _{2}+\tau _{3},\tau _{1}^0-\tau _{1}}\sum _{m_{s}'} e^{-i(\Lambda _{0}\phi _p -\Lambda _{0}'\phi _{\pi })} C\left( \frac{1}{2}\frac{1}{2}1;m_{s}'m_{s1}\right) \nonumber\\
 &  & \times \sum _{l}C\left( l11;M_{d}-m_{s}'-m_{s1},m_{s}'+m_{s1}\right) Y_{l,M_{d}-m_{s}'-m_{s1}}(\hat{{\ffat \pi }}')\psi _{l}(\pi ') \nonumber\\
&  & \times \sum _{S\pi t}\left( 1-\eta _{\pi }(-)^{S+t}\right) C\left( \frac{1}{2}\frac{1}{2}t;\tau _{2}\tau _{3}\right) C\left( \frac{1}{2}\frac{1}{2}t;\tau _{1}^0,-\tau _{1}\right) \nonumber\\
 &  & \times C\left( \frac{1}{2}\frac{1}{2}S;m_{s2}m_{s3}\Lambda _{0}\right) C\left( \frac{1}{2}\frac{1}{2}S;m_{s1}^0m_{s}'\Lambda _{0}'\right) \nonumber\\
 &  & \times \sum _{\Lambda \Lambda '}d^{S}_{\Lambda _{0}\Lambda }(\theta _p)d^{S}_{\Lambda _{0}'\Lambda '}(\theta _{\pi })T_{\Lambda \Lambda '}^{\pi St}({\bf p},{\ffat \pi };E_p) \label{nd16}. 
\end{eqnarray}
Now let us concentrate on the NN T-matrix elements $T_{\Lambda \Lambda '}^{\pi St}({\bf p},{\bf p}';E_p)$ 
in the momentum-helicity basis. For the calculation of  NN scattering it is convenient to 
choose the z-axis as the direction of the initial momenta ${\bf p}'$. It is shown in Ref.~\cite{nn3d} that 
in this case the azimuthal dependencies of the T-matrix elements can be separated as 
\begin{equation}\label{nn1}
T_{\Lambda \Lambda '}^{\pi St}\left( {\bf p},p'\hat{{\bf z}};E_p\right) = e^{i\Lambda '\phi }T_{\Lambda \Lambda '}^{\pi St}(p,p',\cos \theta ;E_p) ,
\end{equation}
which then allowed to reduce the LSE's for $T_{\Lambda \Lambda '}^{\pi St}\left( {\bf p},p'\hat{{\bf
z}};E_p\right)$ to be the ones for $T_{\Lambda \Lambda '}^{\pi St}(p,p',\cos \theta ;E_p) $. 
Thus, in order to calculate $U_{0}^{(1)}({\bf p},{\bf q})$ we have to find a relation between 
$T_{\Lambda \Lambda '}^{\pi St}({\bf p},{\bf p}';E_p)$ with arbitrary $\hat{{\bf p}}'$ and 
$T_{\Lambda \Lambda '}^{\pi St}(p,p',\cos \theta '';E_p) $, where $\theta ''$ now depends on 
$\hat{{\bf p}}$ and $\hat{{\bf p}}'$. This is done in the following way. First we rotate 
$ T_{\Lambda \Lambda '}^{\pi St}({\bf p},{\bf p}';E_p) $ so that $\hat{{\bf p}}'$ points in the z-direction and then apply the relation given in Eq.~(\ref{nn1}). 
With $R(\hat{{\bf p}}')$ being a rotation operator working in momentum and spin space, it follows according
to  Ref.~\cite{nn3d} that
\begin{eqnarray}
T_{\Lambda \Lambda '}^{\pi St}\left( {\bf p},{\bf p}';E_p\right)  & = & \left. ^{^{}}\right. ^{\pi a}\left\langle {\bf p};\hat{{\bf p}}S\Lambda ;t\right| T(E_p)\left| {\bf p}';\hat{{\bf p}}'S\Lambda ';t\right\rangle ^{\pi a}\nonumber\\
 & = & \left. ^{^{}}\right. ^{\pi a}\left\langle {\bf p};\hat{{\bf p}}S\Lambda ;t\right| T(E_p)R(\hat{{\bf p}}')\left| p' \hat{{\bf z}};\hat{{\bf z}}S\Lambda ';t\right\rangle ^{\pi a}\nonumber\\
 & = & \left. ^{^{}}\right. ^{\pi a}\left\langle {\bf p};\hat{{\bf p}}S\Lambda ;t\right| R(\hat{{\bf p}}')T(E_p)\left| p' \hat{{\bf z}};\hat{{\bf z}}S\Lambda ';t\right\rangle ^{\pi a}\nonumber\\
 & = & \left. ^{^{^{}}}\right. ^{\pi a}\left\langle R^{\dagger }(\hat{{\bf p}}'){\bf p};\hat{{\bf p}}S\Lambda 
;t\left| T(E_p)\right| p' \hat{{\bf z}};\hat{{\bf z}}S\Lambda ';t\right\rangle ^{\pi a} . \label{ndnn1}
\end{eqnarray}
Here 
\begin{equation}
\left| {\bf p};\hat{{\bf p}}S\Lambda ;t \right\rangle ^{\pi a} \equiv \frac{1}{2}\left(1-\eta _{\pi }(-)^{S+t}\right)\left(1+\eta _{\pi }\right) | t \rangle \left| {\bf p};\hat{{\bf p}}S\Lambda \right\rangle ,
\end{equation}
and thus, 
\begin{equation}
R^{\dagger }(\hat{{\bf p}}')\left| {\bf p};\hat{{\bf p}}S\Lambda ;t \right\rangle ^{\pi a} = \frac{1}{2}\left(1-\eta _{\pi }(-)^{S+t}\right)\left(1+\eta _{\pi }\right) | t \rangle R^{\dagger }(\hat{{\bf p}}')\left| {\bf p};\hat{{\bf p}}S\Lambda \right\rangle .
\end{equation}
The action of $R^{\dagger }(\hat{{\bf p}}')$ on the state $ \left| {\bf p};\hat{p}S\Lambda \right\rangle  $ leads to two successive rotations as 
\begin{equation}
R^{\dagger }(\hat{{\bf p}}')\left| {\bf p};\hat{p}S\Lambda \right\rangle = R(0,-\theta ',-\phi ')R(\phi \theta 0)\left| p\hat{{\bf z}};\hat{{\bf z}}S\Lambda \right\rangle ,
\end{equation}
and the result is (see Appendix \ref{TWOROT} for the derivation) given by 
\begin{eqnarray}
R^{\dagger }(\hat{{\bf p}}')\left| {\bf p};\hat{p}S\Lambda \right\rangle & = & e^{i\Lambda \Omega }R(\phi ''\theta ''0)\left| p\hat{{\bf z}};\hat{{\bf z}}S\Lambda \right\rangle \nonumber\\
 & = & e^{i\Lambda \Omega }\left| {\bf p}'';\hat{{\bf p}}''S\Lambda \right\rangle , \label{ndnn2}
\end{eqnarray}
with 
\begin{eqnarray}
\cos \theta '' & = & \cos \theta \cos \theta '+\sin \theta \sin \theta '\cos (\phi -\phi ') \\
\sin \theta ''e^{i\phi ''} & = & -\cos \theta \sin \theta '+\sin \theta \cos \theta '\cos (\phi -\phi ') + i \sin \theta \sin (\phi -\phi ') \\
e^{i\Lambda \Omega } & = & \frac{\sum ^{S}_{N=-S}D^{S\ast }_{N\Lambda '}(\phi '\theta '0)D^{S}_{N\Lambda }(\phi \theta 0)}{D^{S}_{\Lambda '\Lambda }(\phi ''\theta ''0)} ,
\end{eqnarray}
where  $D^{S}_{\Lambda '\Lambda }(\phi \theta 0)$ are the Wigner D-function \cite{rose}. 
Inserting Eq.~(\ref{ndnn2}) and then Eq.~(\ref{nn1}) into Eq.~(\ref{ndnn1}) yields 
\begin{eqnarray}
T_{\Lambda \Lambda '}^{\pi St}({\bf p},{\bf p}';E_p) & = & e^{-i\Lambda \Omega } \left. ^{^{}}\right. ^{\pi a}\left\langle {\bf p}'';\hat{{\bf p}}''S\Lambda ;t\right| T(E_p)\left| p' \hat{{\bf z}};\hat{{\bf z}}S\Lambda ';t\right\rangle ^{\pi a} \nonumber\\
& = & e^{-i\Lambda \Omega }T_{\Lambda \Lambda '}^{\pi St}\left( {\bf p}'',p'\hat{{\bf z}};E_p\right) \nonumber\\
& = & e^{i(\Lambda '\phi ''-\Lambda \Omega )}T_{\Lambda \Lambda '}^{\pi St}(p,p',\cos \theta '';E_p) , \label{nd15} 
\end{eqnarray}
where the exponential factor $ e^{i(\Lambda '\phi ''-\Lambda \Omega )} $ is calculated as
\begin{eqnarray}
e^{i(\Lambda '\phi ''-\Lambda \Omega )} & = & e^{i\Lambda '\phi ''}\frac{\sum ^{S}_{N=-S}D^{S\ast }_{N\Lambda }(\phi \theta 0)D^{S}_{N\Lambda '}(\phi '\theta '0)}{D^{S\ast }_{\Lambda '\Lambda }(\phi ''\theta ''0)}\nonumber \nonumber\\
 & = & \frac{\sum ^{S}_{N=-S}e^{iN(\phi -\phi ')}d^{S}_{N\Lambda }(\theta )d^{S}_{N\Lambda '}(\theta ')}{d^{S}_{\Lambda '\Lambda }(\theta '')}.
\end{eqnarray}
Returning to Eq.~(\ref{nd16}), by means of the relation given in Eq.~(\ref{nd15}) we arrive at our final expression for $U_0^{(1)}({\bf p},{\bf q})$: 
\begin{eqnarray}
U_{0}^{(1)}({\bf p},{\bf q}) & = & \frac{(-)^{\frac{1}{2}+\tau _{1}}}{4\sqrt{2}}\delta _{\tau _{2}+\tau _{3},\tau _{1}^0-\tau _{1}}\sum _{m_{s}'} e^{-i(\Lambda _{0}\phi _p -\Lambda _{0}'\phi _{\pi })} C\left( \frac{1}{2}\frac{1}{2}1;m_{s}'m_{s1}\right) \nonumber\\
 &  & \times \sum _{l}C\left( l11;M_{d}-m_{s}'-m_{s1},m_{s}'+m_{s1}\right) Y_{l,M_{d}-m_{s}'-m_{s1}}(\hat{{\ffat \pi }}')\psi _{l}(\pi ') \nonumber\\
&  & \times \sum _{S\pi t}\left( 1-\eta _{\pi }(-)^{S+t}\right) C\left( \frac{1}{2}\frac{1}{2}t;\tau _{2}\tau _{3}\right) C\left( \frac{1}{2}\frac{1}{2}t;\tau _{1}^0,-\tau _{1}\right) \nonumber\\
&  & \times C\left( \frac{1}{2}\frac{1}{2}S;m_{s2}m_{s3}\Lambda _{0}\right) C\left( \frac{1}{2}\frac{1}{2}S;m_{s1}^0m_{s}'\Lambda _{0}'\right) \nonumber\\
 &  & \times \sum _{\Lambda \Lambda '}d^{S}_{\Lambda _{0}\Lambda }(\theta _p)d^{S}_{\Lambda _{0}'\Lambda '}(\theta _{\pi }) e^{i(\Lambda '\phi '-\Lambda \Omega )}T_{\Lambda \Lambda '}^{\pi St}(p,\pi,\cos \theta ';E_p) \label{nd54},
\end{eqnarray}
with
\begin{eqnarray}
\cos \theta ' & = & \cos \theta _p\cos \theta _{\pi }+\sin \theta _p\sin \theta _{\pi }\cos (\phi _p-\phi _{\pi })\\
e^{i(\Lambda '\phi '-\Lambda \Omega )} & = & \frac{\sum ^{S}_{N=-S}e^{iN(\phi _p-\phi _{\pi })}d^{S}_{N\Lambda }(\theta _p)d^{S}_{N\Lambda '}(\theta _{\pi })}{d^{S}_{\Lambda '\Lambda }(\theta ')} .
\end{eqnarray}


\section{Relativistic Kinematics in the Nd Break-Up Amplitude}

In the previous section the break-up operator $U_{0}({\bf p},{\bf q})$ is derived within the 
framework of the nonrelativistic Faddeev theory. Since our goal is to study break-up reactions at
intermediate energies, we want to consider the influence of relativistic kinematics. This means, that we not
only have to employ relativistic energy-momentum relations, 
but more importantly have to reevaluate the Jacobi momenta,  carry out corresponding
Lorentz transformations to the two- and three-particle c.m. subsystems, 
and employ a relativistic description of the
cross section. 
For our derivation we adopt the formulation given in  Ref.~\cite{fong}. For clarity we will describe the
most important steps in detail.

\subsection{Jacobi Momenta}

Let a system be described by the energy and momentum vector $(E,{\bf k})$ in one frame. Then, in 
a different frame moving with  relative velocity ${\bf u}$, the system is described by $(E',{\bf k}')$,
connected by a Lorentz transformation $L({\bf u})$ to the first frame,
\begin{eqnarray}
(E',{\bf k}') & \equiv & L({\bf u})(E,{\bf k}) \label{nd20} \\
{\bf k}' & = & {\bf k}+(\gamma -1)({\bf k}\cdot \hat{{\bf u}})\hat{{\bf u}}-\gamma E{\bf u} \label{nd21}\\
E' & = & \gamma (E-{\bf k}\cdot {\bf u}) \label{nd37}\\
\gamma & \equiv & \frac{1}{\sqrt{1-u^{2}}} \label{nd22} .
\end{eqnarray}
Using these relations we can bring our 3N system from the laboratory frame to the c.m.~frame and find 
the corresponding Jacobi momenta ${\bf p}$ and ${\bf q}$ in the final state and ${\bf q}_0$ in the initial state. 

As in the previous section we choose without loss of generality the two-nucleon subsystem to consist of
nucleons 2 and 3, and let nucleon 1 be the spectator. 
To derive the relative momentum ${\bf p}$ of the subsystem in its c.m.~frame, we employ the Lorentz transformation
\begin{eqnarray}
{\bf u} & = & \frac{{\bf k}_{23}}{E_{23}}\label{nd27}\\
(E_{2}',{\bf p}) & \equiv & (E_{2}',{\bf k}_{2}') = L({\bf u})(E_{2},{\bf k}_{2})\\
(E_{2}',-{\bf p}) & \equiv & (E_{3}',{\bf k}_{3}') = L({\bf u})(E_{3},{\bf k}_{3}) \label{nd28} .
\end{eqnarray}
Let us define the following quantities for the 23-subsystem,
\begin{eqnarray}
{\bf k}_{23} & \equiv & {\bf k}_{2}+{\bf k}_{3} \label{nd23}\\
E_{23} & \equiv & E_{2}+E_{3} \label{nd24},
\end{eqnarray}
which are connected by a Lorentz invariant relation as 
\begin{equation}
E_{23}^{2}-{\bf k}_{23}^{2} \equiv M_{23}^{2} \geq 4m^2 \label{nd19} .
\end{equation}
Here $M_{23}$ is called the invariant mass of the 23-subsystem and equals the total energy of the 
23-subsystem in its c.m.~frame. According to Eq.~(\ref{nd21}) and the transformations given in 
Eqs.~(\ref{nd27})-(\ref{nd28}), the Jacobi momentum ${\bf p}$ is given as \cite{costerlee}
\begin{eqnarray}
{\bf p} & = & \frac{1}{2}({\bf k}_{2}-{\bf k}_{3})-\frac{1}{2}{\bf k}_{23}\left( \frac{E_{2}-E_{3}}{E_{23}+M_{23}}\right) \label{nd25}.
\end{eqnarray}
The last term in Eq.~(\ref{nd25})
exhibits the relativistic effect in the definition of the relative momentum. 
Since in the c.m.~frame 
 the energy $E_2'$ is equal to $E_3'$,  the total energy $M_{23}$ in the 23-frame is given as 
\begin{equation}
M_{23} = E_{2}' + E_{3}' = 2 E_{2}' = 2\sqrt{m^{2}+{k_{2}'}^{2}} = 2\sqrt{m^{2}+p^{2}} \label{nd26}.
\end{equation}
Starting with Eq.~(\ref{nd26}), employing energy and momentum conservation together with 
Eqs.~(\ref{nd23})-(\ref{nd19}) the magnitude of ${\bf p}$ is calculated as 
\begin{eqnarray}
p & = & \frac{1}{2}\sqrt{m_{d}^{2}-2m^{2}+2m_{d}(E_{lab}-E_{1})-2E_{lab}E_{1}+2k_{lab}k_{1}\cos \theta _{lab}} \label{nd38}.
\end{eqnarray}
Knowing the total energy $M_{23}$  in the 23-frame one can calculate the kinetic energy $E_p$ in the
 23-subsystem, that is 
\begin{equation}\label{ndkin}
E_p = M_{23} - 2m = 2\sqrt{m^2 + p^2} - 2m . 
\end{equation}
Thus the NN T-matrix elements $T_{\Lambda \Lambda '}^{\pi St}(p,\pi,\cos \theta ';E_p) $ in Eq.~(\ref{nd54})
will be  calculated for the energy $E_p$ given in Eq.~(\ref{ndkin}). 

In order to obtain the Jacobi momentum ${\bf q}$ we apply the following Lorentz transformation 
\begin{eqnarray}
{\bf u} & = & \frac{{\bf k}_{lab}}{E_{0}}\\
(E_{1}',{\bf q}) & \equiv & (E_{1}',{\bf k}_{1}') = L({\bf u})(E_{1},{\bf k}_{1}) \label{ndranu}\\
(E_{23}',-{\bf q}) & \equiv & (E_{23}',{\bf k}_{23}') = L({\bf u})(E_{23},{\bf k}_{23}) ,
\end{eqnarray}
where $E_0$ is the total energy in the laboratory frame,
\begin{equation}
E_0 \equiv m_d + E_{lab} = E_1 + E_{23} \label{nd33} . 
\end{equation}
Similar to ${\bf p}$, the Jacobi momentum
${\bf q}$ acquires a relativistic correction term and is given by
\begin{equation}
{\bf q} = \frac{1}{2}({\bf k}_{1}-{\bf k}_{23})+\frac{{\bf k}_{lab}}{2M_{0}}\left( \frac{({\bf k}_{1}-{\bf k}_{23})\cdot {\bf k}_{lab}}{E_{0}+M_{0}}-(E_{1}-E_{23})\right) \label{nd36} .
\end{equation}
Here $M_0$ is c.m. total energy or the invariant mass of the 3N system
\begin{equation}\label{nd34}
M_0 \equiv E_1' + E_{23}' ,
\end{equation}
which is connected to the laboratory total energy $E_0$ and the laboratory total momentum 
${\bf k}_{lab}$ by the following Lorentz invariant relation
\begin{equation}\label{nd34b}
E_0^2 - {\bf k}_{lab}^2 = M_0^2 \geq 9m^2 .
\end{equation}
The energies $E_1'$ and $E_{23}'$ are given in terms of the magnitudes of Jacobi momenta ${\bf p}$ and ${\bf q}$ as
\begin{eqnarray}
E_{1}' & = & \sqrt{m^{2}+{k_{1}'}^{2}}=\sqrt{m^{2}+q^{2}}\label{nd31}\\
E_{23}' & = & \sqrt{M_{23}^{2}+{k_{23}'}^{2}}=\sqrt{4(m^{2}+p^{2})+q^{2}} . \label{nd31b}
\end{eqnarray}
With Eqs.~(\ref{ndranu})-(\ref{nd33}), (\ref{nd34}) and (\ref{nd31})-(\ref{nd31b}) we get 
for the scalar product in Eq.~(\ref{nd36})
\begin{eqnarray}
({\bf k}_{1}-{\bf k}_{23})\cdot {\bf k}_{lab} & = & k_{1}^{2}-k_{23}^{2}\nonumber\\
 & = & E_{1}^{2}-E_{23}^{2}-m^{2}+M_{23}^{2}\nonumber\\
 & = & (E_{1}-E_{23})E_{0}-({E_{1}'}^{2}-{E_{23}'}^{2})\nonumber\\
 & = & (E_{1}-E_{23})E_{0}-(E_{1}'-E_{23}')M_{0} \label{nd35}.
\end{eqnarray}
Inserting this result into Eq.~(\ref{nd36}) leads to a final expression for ${\bf q}$, 
\begin{eqnarray}
{\bf q} & = & \frac{1}{2}({\bf k}_{1}-{\bf k}_{23})+\frac{{\bf k}_{lab}}{2M_{0}}\left( \frac{(E_{1}-E_{23})E_{0}-(E_{1}'-E_{23}')M_{0}}{E_{0}+M_{0}}-(E_{1}-E_{23})\right) \nonumber\\
& = & \frac{1}{2}({\bf k}_{1}-{\bf k}_{23})-\frac{1}{2}{\bf k}_{lab}\left( \frac{E_{1}-E_{23}+E_{1}'-E_{23}'}{E_{0}+M_{0}}\right) \nonumber\\
& = & {\bf k}_{1}-{\bf k}_{lab}\left( \frac{E_{1}+E_{1}'}{E_{0}+M_{0}}\right) \label{nd41}.
\end{eqnarray}
For the last equality total momentum conservation was employed. With the help of Eqs.~(\ref{nd31}) 
and (\ref{nd31b}) we can write $M_0$ as 
\begin{equation}
M_{0} = E_{1}' + E_{23}' = \sqrt{m^{2}+q^{2}}+\sqrt{4(m^{2}+p^{2})+q^{2}} \label{nd32}. 
\end{equation}
Starting from Eq.~(\ref{nd32}) we obtain after some algebra the magnitude of ${\bf q}$ as 
\begin{equation}
q = \frac{1}{2 M_{0}}\sqrt{\left\{M_{0}^{2} - (5m^{2}+4p^{2})\right\}^{2}-16m^{2}(m^{2}+p^{2})} \label{nd39} ,
\end{equation}
where the value of $M_0$ can be calculated from Eq.~(\ref{nd34b}) as 
\begin{eqnarray}
M_0 & = & \sqrt{E_{0}^{2}-k_{lab}^{2}} = \sqrt{m^{2}+m_{d}^{2}+2m_{d}E_{lab}} \label{nd40} .
\end{eqnarray}

Finally we turn to the initial state, calculate the Jacobi momentum ${\bf q}_0$, the energies $E_{lab}'$ of 
the incoming nucleon and $E_d'$ of the deuteron in the c.m.~frame. Replacing
 in Eq.~(\ref{nd36}) the quantities ${\bf k}_1$ with ${\bf k}_{lab}$, $E_1$ with $E_{lab}$, 
${\bf k}_{23}$ with zero and $E_{23}$ with $m_d$, the Jacobi momentum ${\bf q}_0$ is given as  
\begin{eqnarray}
{\bf q}_{0} & = & \frac{1}{2}{\bf k}_{lab}+\frac{{\bf k}_{lab}}{2M_{0}}\left( \frac{k_{lab}^{2}}{E_{0}+M_{0}}-(E_{lab}-m_{d})\right) \nonumber\\
 & = & \frac{1}{2}{\bf k}_{lab}+\frac{{\bf k}_{lab}}{2M_{0}}\left( E_{0}-M_{0}-(E_{lab}-m_{d})\right) \nonumber\\
 & = & \frac{m_{d}}{M_{0}}{\bf k}_{lab} \label{nd42}.
\end{eqnarray}
The energies $E_{lab}'$ and $E_d'$ are obtained using Eq.~(\ref{nd37}) as
\begin{eqnarray}
E_{lab}' & = & \frac{E_{0}}{M_{0}}\left(E_{lab}-\frac{k_{lab}^{2}}{E_{0}}\right) = \frac{1}{M_{0}}(m^{2}+m_{d}E_{lab})\\
E_{d}' & = & \frac{E_{0}}{M_{0}}m_{d} = \frac{1}{M_0}(E_{lab}+m_{d})m_{d} .
\end{eqnarray}
It can be shown that $E_{lab}'$ and $E_d'$ sum up to $M_0$ as required by total energy conservation in the c.m. frame. 

\subsection{S-Matrix}

For a correct treatment of relativistic kinematics, we still need to connect the S-matrix element in the
laboratory frame to the Nd break-up amplitude. Suppressing all discrete quantum numbers we define the
S-matrix element for the break-up process in the laboratory frame as
\begin{equation}\label{nd61b}
S({\bf k}_1,{\bf k}_2,{\bf k}_3) \equiv \langle {\bf k}_1{\bf k}_2{\bf k}_3 |S| {\bf k}_{lab}{\bf k}_d\rangle
\end{equation}
and in the c.m. frame as 
\begin{equation}\label{nd61}
S({\bf p},{\bf q}) \equiv \langle {\bf p}{\bf q} |S| {\bf q}_0\rangle 
\equiv \langle {\bf p}{\bf q} |S| {\bf q}_0\rangle .
\end{equation}
In the above equation we omitted the deuteron state $|\Psi_d\rangle$ in the initial state, but showed
its momentum ${\bf k}_d$ 
in  Eq.~(\ref{nd61b}), though its value is zero. The momenta ${\bf p}$, ${\bf q}$, and ${\bf q_0}$ are
the Jacobi momenta calculated in Eqs.~(\ref{nd25}), (\ref{nd41}), and (\ref{nd42}). 
The states $| {\bf k}_1{\bf k}_2{\bf k}_3 \rangle$ and $| {\bf k}_{lab}{\bf k}_d \rangle$  are 
related to the states $| {\bf p}{\bf q} \rangle$ and  $| {\bf q}_0 \rangle$ by the following relations,
\begin{eqnarray}
| {\bf k}_1{\bf k}_2{\bf k}_3 \rangle & \equiv & | {\bf k}_1\rangle |{\bf k}_2{\bf k}_3 \rangle \nonumber\\
& = & J^{-\frac{1}{2}}({\bf k}_2,{\bf k}_3)| {\bf k}_1\rangle | {\bf p}{\bf k}_{23} \rangle \nonumber\\
& \equiv & J^{-\frac{1}{2}}({\bf k}_2,{\bf k}_3)J^{-\frac{1}{2}}({\bf k}_1,{\bf k}_{23}) | {\bf p}{\bf q}\rangle |{\bf k}_1+{\bf k}_{23} \rangle \label{nd43}\\
| {\bf k}_{lab}{\bf k}_d \rangle & \equiv & J^{-\frac{1}{2}}({\bf k}_{lab},{\bf k}_d)| {\bf q}_0 \rangle |{\bf k}_{lab} \rangle \label{nd44}.
\end{eqnarray}
The Jacobian $J({\bf k}_2,{\bf k}_3)$ of the transformation from the variables $({\bf k}_2,{\bf k}_3)$ to 
$({\bf p},{\bf k}_{23})$ is given by \cite{fong} 
\begin{equation}
J({\bf k}_2,{\bf k}_3) = \left|\frac{\partial({\bf k}_2,{\bf k}_3)}{\partial({\bf p},{\bf k}_{23})} \right| = \frac{E_2 E_3}{E_{23}}\frac{M_{23}}{E_2' E_3'} = \frac{4 E_2 E_3}{E_{23}M_{23}} \label{nd51},
\end{equation}
where the last equality results by means of Eq.~(\ref{nd26}) for $M_{23}$. Similarly, 
the Jacobians $J({\bf k}_1,{\bf k}_{23})$ in Eq.~(\ref{nd43}) and $J({\bf k}_{lab},{\bf k}_d)$ in 
Eq.~(\ref{nd44}) are given as
\begin{eqnarray}
J({\bf k}_1,{\bf k}_{23}) & = & \left|\frac{\partial({\bf k}_1,{\bf k}_{23})}{\partial({\bf q},{\bf k}_1+{\bf k}_{23})} \right| = \frac{E_1 E_{23}}{E_0}\frac{M_0}{E_1' E_{23}'} \\
J({\bf k}_{lab},{\bf k}_d) & = & \left|\frac{\partial({\bf k}_{lab},{\bf k}_d)}{\partial({\bf q}_0,{\bf k}_{lab})} \right| = \frac{E_{lab} m_d}{E_0}\frac{M_0}{E_{lab}' E_d'} .
\end{eqnarray}
Thus, one can finally relate  $S({\bf k}_1,{\bf k}_2,{\bf k}_3) $ to $S({\bf p},{\bf q}) $ as 
\begin{eqnarray}
S({\bf k}_1,{\bf k}_2,{\bf k}_3) & = & \langle {\bf k}_1{\bf k}_2{\bf k}_3 |S| {\bf k}_{lab}{\bf k}_d\rangle \nonumber\\
& = & J^{-\frac{1}{2}}({\bf k}_2,{\bf k}_3)J^{-\frac{1}{2}}({\bf k}_1,{\bf k}_{23}) J^{-\frac{1}{2}}({\bf k}_{lab},{\bf k}_d) \langle {\bf k}_1+{\bf k}_{23} |\langle {\bf p}{\bf q} |S| {\bf q}_0\rangle |{\bf k}_{lab} \rangle \nonumber \\
& = & \delta ({\bf k}_1+{\bf k}_{23} - {\bf k}_{lab}) \left\{J({\bf k}_2,{\bf k}_3)J({\bf k}_1,{\bf k}_{23}) J({\bf k}_{lab},{\bf k}_d) \right\}^{-\frac{1}{2}} S({\bf p},{\bf q}) \label{nd45},
\end{eqnarray}
where the delta function ensures total momentum conservation. 

As last step we need to connect $S({\bf k}_1,{\bf k}_2,{\bf k}_3) $ to the break-up amplitude 
$U_0({\bf p},{\bf q}) $ defined in Eq.~(\ref{nd7}), and which is related to $S({\bf p},{\bf q})$ as 
\begin{equation}\label{nd46}
S({\bf p},{\bf q}) = -2\pi i \delta (E_1' + E_{23}' - E_{lab}' - E_{d}') U_0({\bf p},{\bf q}) . 
\end{equation}
Inserting Eq.~(\ref{nd46}) into Eq.~(\ref{nd45}) this gives 
\begin{eqnarray}
S({\bf k}_1,{\bf k}_2,{\bf k}_3) & = & - 2\pi i \delta ({\bf k}_1+{\bf k}_{23} - {\bf k}_{lab}) \delta (E_1' + E_{23}' - E_{lab}' - E_{d}') \nonumber\\
&  & \times \left\{J({\bf k}_2,{\bf k}_3)J({\bf k}_1,{\bf k}_{23}) J({\bf k}_{lab},{\bf k}_d) \right\}^{-\frac{1}{2}} U_0({\bf p},{\bf q}) \label{nd47}.
\end{eqnarray}
By means of the identity
\begin{eqnarray}
\lefteqn{\delta ({\bf k}_1+{\bf k}_{23} - {\bf k}_{lab}) \delta (E_1' + E_{23}' - E_{lab}' - E_{d}')} & & \nonumber\\
& & \qquad \qquad \qquad = \delta ({\bf k}_1+{\bf k}_{23} - {\bf k}_{lab}) \delta (E_1 + E_{23} - E_{lab} - m_{d}) \frac{E_1' + E_{23}' + E_{lab}' + E_{d}'}{E_1 + E_{23} + E_{lab} + m_{d}} , \label{nd48} 
\end{eqnarray}
we obtain the relation between $S({\bf k}_1,{\bf k}_2,{\bf k}_3) $ and $U_0({\bf p},{\bf q})$ as 
\begin{eqnarray}
S({\bf k}_1,{\bf k}_2,{\bf k}_3) & = & - 2\pi i \delta ({\bf k}_1+{\bf k}_{23} - {\bf k}_{lab}) \delta (E_1 + E_{23} - E_{lab} - m_{d}) \nonumber\\
&  & \times \frac{E_1' + E_{23}' + E_{lab}' + E_{d}'}{E_1 + E_{23} + E_{lab} + m_{d}}\left\{J({\bf k}_2,{\bf k}_3)J({\bf k}_1,{\bf k}_{23}) J({\bf k}_{lab},{\bf k}_d) \right\}^{-\frac{1}{2}} U_0({\bf p},{\bf q}) \nonumber\\
& = & - 2\pi i \delta ({\bf k}_1+{\bf k}_{23} - {\bf k}_{lab}) \delta (E_1 + E_{23} - E_{lab} - m_{d}) \Gamma ({\bf p},{\bf q}) U_0({\bf p},{\bf q}) \label{nd49} .
\end{eqnarray}
Here the function $\Gamma ({\bf p},{\bf q}) $ is defined as 
\begin{eqnarray}
\Gamma ({\bf p},{\bf q}) & \equiv & \frac{E_1' + E_{23}' + E_{lab}' + E_{d}'}{E_1 + E_{23} + E_{lab} + m_{d}}\left\{J({\bf k}_2,{\bf k}_3)J({\bf k}_1,{\bf k}_{23}) J({\bf k}_{lab},{\bf k}_d) \right\}^{-\frac{1}{2}} \nonumber\\
& = & \sqrt{\frac{M_{23} E_1' E_{23}' E_{lab}' E_d'}{4 E_1 E_2 E_3 E_{lab} m_d}} .
\end{eqnarray}
In the nonrelativistic limit the function $\Gamma ({\bf p},{\bf q}) $ equals 1.


\section{Observables in the Proton-Neutron Charge Exchange Reaction}

So far we derived the break-up amplitude $U_0({\bf p},{\bf q})$ in first order in its most general form.
As application of a proton-deuteron break-up  process we consider the (p,n) charge exchange reaction. 
In the experiments we are going to analyze only the scattered neutron is detected. Thus, when 
calculating the observables of this reaction, all possible directions of the two undetected 
protons must be taken into account. This is accomplished numerically by integrating over the 
relative direction $\hat{{\bf p}}$ between the two protons. In our numerical application, we consider
 the spin averaged differential cross section in the $d(p,n)pp$ reaction and selected  spin observables. 
These are the neutron polarization $P_0$ in the $d(p,\vec{n})pp$ reaction, the analyzing power $A_y$ in the 
$d(\vec{p},n)pp$ reaction and the polarization-transfer coefficients $ D_{ij} $ in the 
$d(\vec{p},\vec{n})pp$ reaction. 
Comprehensive descriptions and derivations of these observables can be found in e.g. Ref.~\cite{1}. 

The nonrelativistic cross section is given as 
\begin{equation} 
\frac{d^5\sigma }{dE_1 d\hat{{\bf k}}_1} = (2\pi )^4\frac{m^3pk_1}{2k_{lab}}\frac{1}{6}\sum_{{m_{s1}m_{s2}m_{s3} \atop m_{s1}^0M_d}}\int d\hat{{\bf p}} |U_0({\bf p},{\bf q})|^2 , \label{nd57}
\end{equation}
where ${\bf k}_1$ determines ${\bf q}$ via Eq.~(\ref{nda1}) and $p$ via Eqs.~(\ref{nd55}) and (\ref{nda1}). 
The relativistic cross section is given by 
\begin{eqnarray}
d\sigma & = & (2 \pi )^4 \frac{E_{lab}}{k_{lab}} \delta ({\bf k}_1+{\bf k}_{23} - {\bf k}_{lab}) \delta (E_1 + E_{23} - E_{lab} - m_{d}) \nonumber\\
 &  & \times \Gamma ^2 ({\bf p},{\bf q}) |U_0({\bf p},{\bf q})|^2 d{\bf k}_1d{\bf k}_2d{\bf k}_3 \label{nd53} ,
\end{eqnarray}
where the S-matrix element from Eq.~(\ref{nd49}) is used, and all energies obey the relativistic
energy-momentum relation. 
The cross section for the inclusive Nd break-up process is calculated as  function of the 
direction $\hat{{\bf k}}_1$ and the kinetic energy $E_{k,1} = E_1 - m$ of the detected nucleon. Using 
the relation
\begin{equation}
d{\bf k}_1 = dk_1 k_1^2 d\hat{{\bf k}}_1 = E_1 k_1 dE_1 d\hat{{\bf k}}_1 = E_1 k_1 dE_{k,1} d\hat{{\bf k}}_1 \label{nd52} 
\end{equation}
leads to 
\begin{eqnarray}
\frac{d\sigma }{dE_1 d\hat{{\bf k}}_1} & = & (2 \pi )^4 \frac{E_{lab} E_1 k_1}{k_{lab}} \nonumber\\
 &  & \times \int d{\bf k}_2d{\bf k}_3 \delta ({\bf k}_1+{\bf k}_{23} - {\bf k}_{lab}) \delta (E_1 + E_{23} - E_{lab} - m_{d}) \Gamma ^2 ({\bf p},{\bf q}) |U_0({\bf p},{\bf q})|^2 \nonumber\\
& = & (2 \pi )^4 \frac{E_{lab} E_1 k_1}{k_{lab}} \nonumber\\
 &  & \times \int d{\bf k}_{23}d{\bf p} J({\bf k}_2,{\bf k}_3) \delta ({\bf k}_1+{\bf k}_{23} - {\bf k}_{lab}) \delta (E_1 + E_{23} - E_{lab} - m_{d}) \Gamma ^2 ({\bf p},{\bf q}) |U_0({\bf p},{\bf q})|^2 \nonumber\\
& = & (2 \pi )^4 \frac{E_{lab} E_1 k_1}{k_{lab}} \int d\hat{{\bf p}} dp \, p^2 \frac{4 E_2 E_3}{E_{23}M_{23}}  \delta (E_1 + E_{23} - E_{lab} - m_{d}) \Gamma ^2 ({\bf p},{\bf q}) |U_0({\bf p},{\bf q})|^2 \nonumber\\
& = & (2 \pi )^4 \frac{E_{lab} E_1 k_1 p}{k_{lab}M_{23}} \int d\hat{{\bf p}} E_2 E_3 \Gamma ^2 ({\bf p},{\bf q}) |U_0({\bf p},{\bf q})|^2 \label{nd50}. 
\end{eqnarray}
In addition we  used Eq.~(\ref{nd51}) for $J({\bf k}_2,{\bf k}_3)$ together with
\begin{eqnarray}
dE_{23} & = & d \sqrt{M_{23}^2 + k_{23}^2} = d \sqrt{4(m^2 + p^2) + k_{23}^2} = \frac{4p}{E_{23}} dp .
\end{eqnarray}
Defining a function $\rho (p,q) $ by
\begin{eqnarray}
\rho (p,q) & \equiv & \frac{2E_{lab} E_1 E_2 E_3 }{M_{23}} \Gamma ^2 ({\bf p},{\bf q}) = \frac{E_1' E_{23}' E_{lab}' E_d'}{2 m_d} ,
\end{eqnarray}
allows to write the relativistic cross section similar to the nonrelativistic one, that is 
\begin{equation}\label{nd57b}
\frac{d\sigma }{dE_1 d\hat{{\bf k}}_1} = (2 \pi )^4 \frac{\rho (p,q) p k_1}{2k_{lab}} \frac{1}{6}\sum_{{m_{s1}m_{s2}m_{s3} \atop m_{s1}^0M_d}}\int d\hat{{\bf p}} \, |U_0({\bf p},{\bf q})|^2 .
\end{equation}
Here we restore the summation over final spins and the averaging over initial spins states. 
In the nonrelativistic case, the function $\rho (p,q) $ reduces to $m^3$, leading to   Eq.~(\ref{nd57}). 
Next we give the polarization $P_0$, the analyzing power $A_y$ and polarization-transfer coefficients 
$ D_{ij}\equiv \frac{1}{6I_{0}}Tr\left\{ U_0({\ffat \sigma }\cdot \hat{{\bf j}})U_0^{\dagger }
({\ffat \sigma }\cdot \hat{{\bf i}})\right\} $ as 
\begin{eqnarray}
P_0 & = & \frac{1}{6I_{0}}Tr\left\{ U_0U_0^{\dagger }({\ffat \sigma }\cdot \hat{{\bf n}})\right\} \label{nd58}\\
A_{y} & = & \frac{1}{6I_{0}}Tr\left\{ U_0({\ffat \sigma }\cdot \hat{{\bf n}})U_0^{\dagger }\right\} \\
D_{n'n} & = & \frac{1}{6I_{0}}Tr\left\{ U_0({\ffat \sigma }\cdot \hat{{\bf n}})U_0^{\dagger }({\ffat \sigma }\cdot \hat{{\bf n}}')\right\} \\
D_{s's} & = & \frac{1}{6I_{0}}Tr\left\{ U_0({\ffat \sigma }\cdot \hat{{\bf s}})U_0^{\dagger }({\ffat \sigma }\cdot \hat{{\bf s}}')\right\} \\
D_{l's} & = & \frac{1}{6I_{0}}Tr\left\{ U_0({\ffat \sigma }\cdot \hat{{\bf s}})U_0^{\dagger }({\ffat \sigma }\cdot \hat{{\bf l}}')\right\} \\
D_{s'l} & = & \frac{1}{6I_{0}}Tr\left\{ U_0({\ffat \sigma }\cdot \hat{{\bf l}})U_0^{\dagger }({\ffat \sigma }\cdot \hat{{\bf s}}')\right\} \\
D_{l'l} & = & \frac{1}{6I_{0}}Tr\left\{ U_0({\ffat \sigma }\cdot \hat{{\bf l}})U_0^{\dagger }({\ffat \sigma }\cdot \hat{{\bf l}}')\right\} ,
\end{eqnarray}
where 
\begin{equation}\label{nd59}
I_0 \equiv \frac{1}{6}Tr\left\{ U_0U_0^{\dagger }\right\} .
\end{equation}
For simplicity we suppressed the $\hat{{\bf p}}$-integration in Eqs.~(\ref{nd58})-(\ref{nd59}). 
The unit vectors $\hat{{\bf n}}$, $\hat{{\bf l}}$, $\hat{{\bf s}}$, $\hat{{\bf n}}'$, $\hat{{\bf l}}'$ and 
$\hat{{\bf s}}'$ are defined as 
\begin{equation}
\hat{{\bf n}} = \hat{{\bf n}}' \equiv  \frac{{\bf k}_{lab} \times {\bf k}_1 }{|{\bf k}_{lab} \times {\bf k}_1 |} , \qquad \hat{{\bf l}} \equiv \hat{{\bf k}}_{lab} , \qquad \hat{{\bf s}} \equiv \hat{{\bf n}} \times \hat{{\bf l}} , \qquad \hat{{\bf l}}' \equiv \hat{{\bf k}}_1 , \qquad \hat{{\bf s}}' \equiv \hat{{\bf n}}' \times \hat{{\bf l}}' .
\end{equation}

For the explicit calculation of the Nd break-up amplitude $U_0({\bf p},{\bf q})$, 
Eq.~(\ref{nd54}), we need the NN T-matrix elements. They are obtained by solving the Lippmann-
Schwinger (LS) equations for a given NN potential in 3D as described in Ref.~\cite{nn3d} at
fixed momenta, angles, and energies. The matrix elements
 $T_{\Lambda \Lambda '}^{\pi St}(p,\pi,\cos \theta ';E_p)$ are then obtained
by interpolating in $\pi$, $\cos \theta '$, and $E_p$. 
This is more economic than solving the LS equation every  time for the corresponding
energies and initial momenta.  Similarly the partial wave deuteron wave function components
are calculated once and interpolated to the  momenta $\pi '$. 
For all the interpolations we use the modified cubic Hermite splines of Ref.~\cite{5}. 
Typical grid sizes for our calculations are 40 points for $E_p$ (and $p$), 80 points for
$\cos \theta '$, and 50 points for  $\pi$. This grid is then used for all our break-up
calculations from 5 to 500 MeV.

\section{Results and Discussion}

Our numerical calculations are based two different NN potentials, the Bonn-B \cite{16} and AV18
\cite{av18} potentials. In the discussion of our results we consider three different aspects. First,
since we present a new way of calculating the break-up process, namely without  partial wave
decomposition, we need to compare our results to those obtained in traditional partial wave based
calculations. After having established the feasibility and correctness of our calculations, we will
address two physical questions, namely the importance of rescattering contributions in the (p,n) charge
exchange reaction at a moderately high energy and the effect of different descriptions of the
kinematics, i.e. we compare the relativistic treatment introduced in Section III to the
nonrelativistic description.

\subsection{Comparison with Partial-Wave Calculations}

In this section we compare our 3D calculations with traditional PW calculations at different
energies. The aim here is twofold. First, we want to convince ourselves that our newly developed 3D
formulations agrees with well established PW calculations \cite{witalapers}. Secondly, we want
to find out from which energy regime on our 3D method surpasses the state-of-the-art PW calculations.  
As already mentioned in the introduction, in PW calculations the number of partial waves necessary for 
a converged result proliferated as the energies increases, leading to limitations with respect to
computational feasibility and accuracy.  Our new 3D scheme does not suffer from these limitations,
the computational effort is in principle independent of the projectile energy considered. 

Our first comparison is carried out at a proton incident energy of $E_{lab} = 100$~MeV 
and a neutron laboratory scattering angle $\theta _{lab} = 13^o$.  The calculations, which are
based on the Bonn-B potential are given in Fig.~\ref{ndrfig1}, which shows the differential cross
section, the analyzing power $A_y$ and the  polarization-transfer coefficient $D_{sl}$.  The solid
lines represent our 3D calculations, the dashed lines the corresponding PW calculations
\cite{witalapers}.  
Here we use the notation $j$ for the highest 2N total angular momentum taken into account in the PW 
calculation, and $J$ for the highest 3N total angular momentum. 
The figure shows that both lines are almost indistinguishable, thus validating our new scheme. 
At this point we also would like to mention, that the channels used in the PW calculation, 
namely $j = 7$ and $J = 31/2$ constitute todays limits for a PW calculation. In addition, we carried
out comparisons at lower energies, e.g. at 16~MeV, where a PW calculation with $j = 5$ and $J = 31/2$
is in perfect agreement with our 3D calculations. 

Next, we turn to a slightly higher projectile energy, $E_{lab} = 197$~MeV, and carry out the same
comparison. The results for the differential cross section, the analyzing power $A_y$ and the
polarization-transfer coefficient $D_{sl}$ are shown in Fig.~\ref{ndrfig2},
where the solid line represents our 3D calculation.
The PW calculations are shown with increasing number of partial waves from $j = 5, \, J = 25/2$ to $j
= 7, \, J = 31/2$. The peak of the differential cross section reveals that each additional 
angular momentum of
the PW calculation results in an additive contribution, 
but even the highest possible number deviates about 9\% from our 3D result.
This is the most extreme case, for the analyzing power $A_y$ and the polarization-transfer
coefficient $D_{sl}$ the PW calculation with $j = 7$ and $J = 31/2$ agrees reasonable well with
our 3D result. It is interesting to note that for $D_{sl}$ the 2N total momentum $j$ is much more
important to reach convergence than the total 3N momentum $J$. 

At this point it is appropriate to  make some general remarks. In this work we restrict our 3D
approach to the leading term in the Faddeev multiple scattering series. Thus we have no insight
whether the fully summed series would lead to a better agreement of 3D and PW approach. 
We also restrict ourselves to semi-exclusive processes, and can not draw any conclusions on 3D and
PW calculations with respect to elastic scattering observables or full break-up observables.
But nevertheless, the comparison in Fig.~\ref{ndrfig2} indicates that around 200~MeV the convergence
of the PW approach has to be carefully checked.

\subsection{Contributions from the Rescattering Terms}

One of the arguments to study the semi-exclusive Nd break-up process in first order in the NN
T-matrix is that at higher energies the rescattering term generated by the solution of the full
Faddeev equations become less important. For a comprehensive study of the importance of those
rescattering terms it would be necessary to compare first order calculations with full Faddeev
calculations over a wide range of projectile energies and for many different experimental
situations. Unfortunately we can not do this at the present stage, since three-dimensional
full Faddeev calculations do not yet exist, and traditional, partial wave based Faddeev
calculations are limited in their energy range. Thus we take as compromise a medium energy of 
about 200 MeV, and 
compare the (p,n) charge exchange observables calculated in our first order 3D approach with
the ones obtained from a full, partial wave based Faddeev calculation. We choose the proton
energy $E_{lab}$=197~MeV, since there exist recent measurements \cite{ndx197}. 

Our calculations are based on two different NN potential models, namely Bonn-B \cite{16} and 
AV18 \cite{av18}. Both potentials are defined below 350~MeV nucleon laboratory energy, which
corresponds to a NN c.m. energy of 175~MeV. In the Nd break-up process in first order the NN
c.m. energy available to the two-nucleon subsystem is fixed in terms of the laboratory momentum
of the final nucleon and the projectile energy. For a projectile energy of about 200~MeV in the
pd scattering process, the maximum NN c.m. energy in the two-body subsystem is about 133 MeV.
Thus, our calculations employs the NN models in an energy regime, where they are perfectly well
defined. Of course, the two potential models exhibit differences in the description of the NN
phase shifts. The model AV18 is one of the socalled high-precision potentials, describing the NN
data base with a  $\chi^2 /datum \approx 1$, whereas Bonn-B has a slightly higher $\chi^2/datum$
value. Thus, there are on-shell differences between those two models, which should lead to
differences in the Nd break-up observables. 

In Figs.~\ref{ndrfig4} and \ref{ndrfig5} we compare the 3D calculations  with PW based full 
Faddeev calculations \cite{witalapers} at 197 MeV proton energy. 
We show the spin averaged differential cross section, the
analyzing power and polarization-transfer coefficients at two angles, 
$\theta _{lab} = 24^o$ and $\theta _{lab} =37^o$ together with experimental results from Ref.~\cite{ndx197}.
The PW full Faddeev calculations use $j = 5,\, J = 31/2$ for the AV18 NN potential, and 
$j = 4,\, J = 31/2$ for the Bonn-B NN potential. Since the solution of the PW full Faddeev 
equations is more involved we restrict ourselves to a lower number of partial waves. However, the
number of partial waves is sufficient to study the importance of rescattering terms in the multiple
scattering series at this energy. The first obvious difference between the two calculations is
the appearance of the final state peak in the differential cross section, which of course is solely due
to rescattering. Furthermore, we see that rescattering contributions have the general tendency to push
the peak of the differential cross section down, though the size of the effect depends on the angle and the
potential. However, the rescattering terms do not affect the position of the peak. We see that the
peak is shifted further away from  the data the larger the neutron scattering angle becomes.
 For both angles, the analyzing power $A_y$  shows the largest effect of rescattering for
small neutron energies, which can be expected, since interactions between outgoing particles should be
larger, when their relative energy is smaller. In both cases,  $A_y$ can only be satisfactorily
described when rescattering terms are taken into account. For the polarization-transfer coefficient
$D_{ll}$ at $\theta _{lab} =24^o$ the situation is similar, rescattering effects are largest for
small neutron energies. For $D_{ss}$ at $\theta _{lab} =37^o$ none of the calculations 
is able to capture the general shape of the data, rescattering effects are visible, but they do not
affect the general shape of the curve as this is the case in the other observables shown in 
Figs.~\ref{ndrfig4} and \ref{ndrfig5}.
From these consideration we have to conclude, that at a projectile energy $E_{lab} \simeq 200$ MeV
rescattering terms still give considerable contributions to 
the full pd break-up amplitude and hence cannot be neglected.
Due to the lack of calculations based on the full Faddeev equations at higher energies, we can not
carry out corresponding studies at higher energies.

\subsection{Relativistic Effects}

In this section we study the effects of relativistic kinematics in the break-up amplitude and follow the
formulation derived in Section III.  
We also want to take full advantage of our 3D formulation and carry out calculations at proton incident
energies higher than 200 MeV, a regime where partial wave based Faddeev calculations become less
competitive. Of course we also realize that the NN potentials from which our NN T-matrix is obtained
are strictly speaking out of their range of validity, i.e.  they do not include important Delta degrees of
freedom. A NN laboratory energy of 350 MeV roughly corresponds to a proton incident laboratory
energy of 260 MeV in the pd scattering process. A comparison of the NN scattering observables with
the calculated ones shows that
even at NN laboratory energies higher than 350 MeV the agreement with data deteriorates relatively
slowly. Nevertheless, this can lead to deficiencies in describing the Nd break-up process at 
$E_{lab} >$260 MeV.  


At first we investigate the effect of relativistic kinematics on the break-up observables
at a low energy, $E_{lab} = 100$ MeV, where it is expected to be small.
In Fig.~\ref{ndrfig7} we show the differential cross section and $A_y $ at $E_{lab} = 100$ MeV 
and a neutron laboratory scattering angle $\theta _{lab} = 24^o$. In the cross section effects are only
visible in the quasi-free-scattering (QFS) peak,
but in general one can say that around 100 MeV relativistic effects are small, and
certainly not the dominant correction to worry about.

Going to a  higher energy, $E_{lab} = 197$ MeV, the relativistic effects increase considerably.  
In Figs.~\ref{ndrfig8} and \ref{ndrfig9} we show the cross section and $A_y$ and two polarization-transfer
coefficients at neutron laboratory scattering angles $\theta _{lab} = 24^o$ and $\theta _{lab} = 37^o$.
Here the QFS peak is visibly enhanced by the use of relativistic kinematics. More importantly, it
location is shifted towards smaller neutron energies, and is now in better agreement with the
experimentally determined peak location. As far as the spin observables are concerned, the relativistic
corrections show the largest effect for the higher neutron energies.  

A proton energy of $E_{lab} = 346$ MeV is the next higher energy at which the (p,n) charge 
exchange reaction is measured \cite{ndx346}. In Fig.~\ref{ndrfig11} we display 
3D calculations with nonrelativistic and relativistic kinematics for a neutron scattering angle 
$\theta _{lab} = 22^o$. Again we observe an increase in the QFS peak and a shift to lower neutron
energies. Here we would like to point out that there is an uncertainty in the data as far as the
location of the QFS peak is concerned. In the experiment there is an uncertainty of the energy, 
at which the pd break-up process exactly occurs. For example, due to the thickness
of the target the proton may have lost some of its energy before it hits and breaks the 
deuteron apart \cite{witalapers}. In this case, the break-up 
the process occurs at an energy slightly different from the calculated one.
The effect of relativistic kinematics on the spin observables is clearly more pronounced compared to
$E_{lab} = 197$ MeV.
We also calculate the break-up process at $E_{lab} = 495$ MeV, though here the uncertainty with respect
to our input NN interactions is largest. In  Fig.~\ref{ndrfig12} we show the differential cross section,
$A_y$ and $D_{ll}$ at $E_{lab} = 495$ MeV for a neutron laboratory scattering angle $\theta _{lab} =
18^o$ together with the measurements from Ref.~\cite{ndx495}.
Here we clearly see that the corrections due to relativistic kinematics push the QFS peak towards
lower neutron energies, and the location of our calculated peak agrees with the measured one.
The effects on the spin observables are now also quite sizable.

With this study we can qualitatively indicate that relativistic effects become important when
going to higher energies. However, we can not make any definite statements, since we only consider
relativistic kinematics. We have not considered effects resulting from boosting the NN T-matrix \cite{kamada}
and Wigner rotations of the spin \cite{polyzou}. Those effects in principle could counterbalance the
kinematic effects. It also remains to be seen, how important rescattering effects will be at those
higher energies. Our calculations based on the first order term and relativistic kinematics
overestimate the differential cross section. That could imply, that rescattering still plays 
an important role at those energies.


\section{Summary and Conclusions}

We formulate and calculate the Nd break-up process based on the Faddeev scheme
 in first order in the multiple scattering expansion
in a three-dimensional fashion which does not rely on any partial wave decomposition. 
The leading term for the Nd break-up amplitude is derived in a representation, 
which uses directly the momentum vectors.
This representation can be connected to 
the momentum-helicity basis, in which we solve for the NN T-matrix in a 3D fashion. 
Special care has to be taken when rotating the NN T-matrix elements,
which occur with arbitrarily oriented initial momenta in the Nd break-up amplitude, 
 such that the NN initial relative 
momenta point into a fixed z-direction. This is needed since 
two nucleon LS equation for
the NN T-matrix is solved in a basis where
the arbitrary z-axis points into the direction of the initial momenta. This leads to an intricate
additional phase factor.

As specific application of our new formulation we calculate the (p,n) charge exchange reaction in the
proton-deuteron break-up process. Here only the outgoing neutron is detected after the break-up. Our
calculations concentrate on spin averaged differential cross sections, neutron polarizations, proton
analyzing powers, and polarization-transfer coefficients at different energies.

First we carry out calculations of observables 
for the leading order term in the NN T-matrix 
at energies which are accessible to traditional,
partial wave based Faddeev calculations. The aim here is twofold. First, we need to establish 
the numerical accuracy and feasibility of our new formulation. 
We establish both by comparing observables calculated in both schemes at a  proton incident energy
$E_{lab}$= 100 MeV, where we find excellent agreement between both calculations.
At $E_{lab}$= 200 MeV we find some slight deviations between the two schemes, especially in the
quasi-free peak of the differential cross section. This can be identified as the onset of a lack of
convergence 
using the typical and feasible number of partial waves
in the traditional partial wave based calculation in that particular observable. From that
we conclude that starting at about 200 MeV, the convergence of partial wave based calculations has to be
checked carefully.

Second, we want to investigate the importance of rescattering terms at a moderately high energy.
Of course, we need full Faddeev calculations here. Since those do not yet exist in a 3D formulation
using realistic NN potentials, we have to resort to partial wave based full Faddeev calculations. 
This of course limits the energy regime we can study. Thus, we compare our calculations at $E_{lab} =
197$ MeV to the PW full Faddeev calculations. We find that at this energy rescattering effects are
still important, and are mostly visible in the cross section and the analyzing power. In addition, we
find that
the PW full Faddeev calculations provide a reasonable description of the (p,n) charge exchange reaction
at 200 MeV. However, we also can detect one obvious deficiency in both schemes,
at larger neutron laboratory
scattering angles the QFS peak is located at slightly too high neutron energies compared to the data.

This leads to the next topic we investigate, namely the effect of relativistic kinematics in the 
Nd break-up reaction. Here we have to employ not only relativistic energy-momentum relations, but also
need to reevaluate the Jacobi momenta by carrying out corresponding Lorentz transformations to the two- and
three-particle c.m. subsystems, and employ a relativistic description of the cross section. 
We compare our 3D calculation based on nonrelativistic kinematics with the corresponding one based on
relativistic kinematics. Though there are no sizable effects at 100 MeV proton incident energy, we find,
that at 200 MeV visible effect occur, mainly the differential cross section. 
Its magnitude increases, but most
importantly, the QFS peak is shifted to the experimentally determined one using relativistic kinematics.
Since our calculations are as easily carried out at 300 or 500 MeV as at 200 MeV, we perform
calculations at $E_{lab}$=346 MeV and 497 MeV, where experimental data are available.
We find that at those higher energies the effects due to the relativistic kinematics are considerably
larger than at  200 MeV. They are now visible not only in the cross section but also in the spin
observables. Even at $E_{lab}$=500 MeV this specific feature prevails, 
namely that the QFS peak is shifted to
lower neutron energies and coincides now with the experimentally determined one. 
However, its magnitude is
larger. With these finding we can qualitatively indicate that relativistic
effects become increasingly important when considering the Nd break-up reaction at higher energies.
However, we have to exercise some caution in the interpretation of our findings, since we have not
considered dynamical relativistic effects, such as boosting of the NN T-matrix. 
It also remains to be seen how important rescattering effects will be in the higher energy regime.

Summarizing, for the first order term in the multiple scattering series in the Faddeev scheme
the 3D approach 
 has proven to be a viable alternative to the established partial wave based calculations. When entering
the intermediate energy regime it may be the 
approach having the most  promise of being successful in
the near future, due to the intrinsic limitations with respect to computational feasibility and
accuracy faced by partial wave calculations at higher energies. It is also clear that the 3D approach,
though having a well defined roadmap ahead, is still facing extensive development needs. The full
Faddeev equations will have to be solved, with the inclusion of three-nucleon forces, which may play
a more dominant role at higher energies. Furthermore, though we consider the effects of
relativistic kinematics,
we have not taken into account the corresponding dynamical effects. And last, but not least,
the underlying input of any 3N calculation, namely the two-nucleon force, is by far less developed
at higher energies than it is for energies below the pion production threshold.

\vfill

\acknowledgements We would like to thank Henryk Wita{\l}a and Jacek Golak for very useful discussions and providing the PW results. This work was performed in part under the
auspices of the Deutsche Akademische Austauschdienst under contract No. A/96/32258, and
the U.~S. Department of Energy under contract No. DE-FG02-93ER40756  with Ohio University. 
We thank the Computer Center of the RWTH  Aachen (Grant P039) for the use of their
facilities. 

\newpage

\appendix

\section{\label{TWOROT}TWO SUCCESSIVE ROTATIONS}

In this appendix we evaluate the rotation of the state $ \left| {\bf p};\hat{p}S\Lambda \right\rangle  $ as
\begin{equation}
R^{\dagger }(\hat{{\bf p}}')\left| {\bf p};\hat{p}S\Lambda \right\rangle = R(0,-\theta ',-\phi ')R(\phi \theta 0)\left| p\hat{{\bf z}};\hat{{\bf z}}S\Lambda \right\rangle . 
\end{equation}
First, we give a few basic definitions and relations required to follow the calculation. 
More details about rotation can be found in e.g. Ref.~\cite{rose}.

A general rotation operator $R(\hat{{\bf p}})$ is defined as 
\begin{equation}
R(\hat{{\bf p}})=R(\phi \theta 0)=e^{-iJ_{z}\phi }e^{-iJ_{y}\theta } ,
\end{equation}
where $ J_{z},\, J_{y} $ are the z- and y-components of the angular momentum operator $ {\bf J} $ and 
$(\theta ,\phi )$ the rotation angles that determine the direction of ${\bf p}$. 
This operator rotates the angular momentum state $|\hat{{\bf z}}jm \rangle $ into
 the state $|\hat{{\bf p}}jm \rangle $,
\begin{equation}
|\hat{{\bf p}}jm \rangle = R(\hat{{\bf p}}) |\hat{{\bf z}}jm \rangle = \sum _{m'} D^j_{m'm}(\hat{{\bf p}}) |\hat{{\bf z}}jm' \rangle , 
\end{equation}
where $D^j_{m'm}(\hat{{\bf p}})$ are the Wigner D-function  defined as 
\begin{equation}
D^j_{m'm}(\hat{{\bf p}}) = D^j_{m'm}(\phi \theta 0) \equiv \langle \hat{{\bf z}}jm'| R(\hat{{\bf p}}) |\hat{{\bf z}}jm \rangle .
\end{equation}
A rotation $R(\alpha \beta \gamma )$ corresponds to a change of the Cartesian coordinates ${\bf r}$ describing the state. The new Cartesian coordinates ${\bf r}'$ are related to the old ones ${\bf r}$ as 
\begin{equation}
{\bf r}' = M(\alpha \beta \gamma ) {\bf r} ,
\end{equation}
where the rotation matrix $M(\alpha \beta \gamma )$ is given as 
\begin{equation}\label{dou8}
M(\alpha \beta \gamma ) = \left(\begin{array}{ccc}
                            \cos \alpha \cos \beta \cos \gamma - \sin \alpha \sin \gamma & 
                            \sin \alpha \cos \beta \cos \gamma + \cos \alpha \sin \gamma & 
                          - \sin \beta  \cos \gamma \\
                          - \cos \alpha \cos \beta \sin \gamma - \sin \alpha \cos \gamma & 
                          - \sin \alpha \cos \beta \sin \gamma + \cos \alpha \cos \gamma & 
                            \sin \beta  \sin \gamma \\
                            \cos \alpha \sin \beta & 
                            \sin \alpha \sin \beta & 
                            \cos \beta
                                 \end{array}\right) .
\end{equation}

\subsection{Two Successive Rotations in Momentum Space}

We denote the rotation operator in momentum space as $R_L(\hat{{\bf p}})$, which is given in term of the orbital angular momentum operator ${\bf L}$ as 
\begin{equation}
R_L(\hat{{\bf p}})=R_L(\phi \theta 0)=e^{-iL_{z}\phi }e^{-iL_{y}\theta } .
\end{equation}

A momentum state $|{\bf p} \rangle $ with $\hat{{\bf p}}$ pointing in the direction
$(\theta ,\phi)$ can be expanded in partial waves as 
\begin{equation}
|{\bf p} \rangle = \sum _{lm} |plm \rangle Y^{\ast}_{lm}(\theta ,\phi) ,
\end{equation}
where $|plm \rangle$ is defined to be quantized along the z-axis. 
The state $|{\bf p} \rangle $ can be obtained by rotating a state $|p \hat{{\bf z}} \rangle $ as follows, 
\begin{eqnarray}
R_L(\hat{{\bf p}}) |p \hat{{\bf z}} \rangle & = & R_L(\phi \theta 0) \sum _{lm} |plm \rangle Y^{\ast}_{lm}(0,0) \nonumber\\
& = & R_L(\phi \theta 0) \sum _{l} |pl0 \rangle \sqrt{\frac{2l + 1}{4\pi }} \nonumber\\
& = & \sum _{l} \sum _{l'm} |pl'm \rangle \langle \hat{{\bf z}}l'm| R_L(\phi \theta 0) |\hat{{\bf z}}l0 \rangle \sqrt{\frac{2l + 1}{4\pi }} \nonumber\\
& = & \sum _{lm} |plm \rangle D^l_{m0}(\phi \theta 0) \sqrt{\frac{2l + 1}{4\pi }} \nonumber\\
& = & \sum _{lm} |plm \rangle Y^{\ast}_{lm}(\theta ,\phi) \nonumber\\
& = & |{\bf p} \rangle . \label{dou4}
\end{eqnarray}
Here we used the relation between the spherical harmonics and the Wigner D-functions,
\begin{equation}
Y^{\ast}_{lm}(\theta ,\phi)= \sqrt{\frac{2l+1}{4\pi }}D^{l}_{m0}(\phi \theta 0) .
\end{equation}
Now we rotate the state $|{\bf p} \rangle $ with an inverse rotation operator $R_L^{-1}(\hat{{\bf p}}') = R_L^{\dagger}(\hat{{\bf p}}') = R_L(0, -\theta ', -\phi ') $. It follows that 
\begin{eqnarray}
R_L^{\dagger}(\hat{{\bf p}}') |{\bf p} \rangle & = & R_L^{\dagger}(\hat{{\bf p}}') \sum _{lm} |plm \rangle Y^{\ast}_{lm}(\theta ,\phi) \nonumber\\
& = & \sum _{lm} \sum _{l'm'} |pl'm' \rangle \langle \hat{{\bf z}}l'm'| R_L^{\dagger}(\hat{{\bf p}}') |\hat{{\bf z}}lm \rangle Y^{\ast}_{lm}(\theta ,\phi) \nonumber\\
& = & \sum _{lm} \sum _{l'm'} |pl'm' \rangle \langle \hat{{\bf z}}l'm'| R_L^{\dagger}(\hat{{\bf p}}') |\hat{{\bf z}}lm \rangle \left\langle \hat{{\bf z}}lm \right. | \hat{{\bf p}} \rangle \nonumber\\
& = & \sum _{l'm'} |pl'm' \rangle \langle \hat{{\bf z}}l'm'| R_L^{\dagger}(\hat{{\bf p}}') | \hat{{\bf p}} \rangle \nonumber\\
& \equiv & \sum _{l'm'} |pl'm' \rangle \langle \hat{{\bf z}}l'm'| \left. \hat{{\bf p}}'' \right\rangle \nonumber\\
& = & \sum _{l'm'} |pl'm' \rangle Y^{\ast}_{l'm'}(\theta '',\phi '') \nonumber\\
& = & R_L(\hat{{\bf p}}'') |p\hat{{\bf z}} \rangle , \label{dou3}
\end{eqnarray}
where we have defined a direction $\hat{{\bf p}}'' $ to be determined by $\hat{{\bf p}} $ and $\hat{{\bf p}}' $ according to 
\begin{equation}
| \hat{{\bf p}}'' \rangle = R^{\dagger}_L(\hat{{\bf p}}') | \hat{{\bf p}} \rangle .
\end{equation}
Inserting Eq.~(\ref{dou4}) into Eq.~(\ref{dou3}) this leads to 
\begin{equation}\label{dou5}
R_L^{\dagger}(\hat{{\bf p}}') R_L(\hat{{\bf p}}) |p \hat{{\bf z}} \rangle = R_L(\hat{{\bf p}}'') |p\hat{{\bf z}} \rangle .
\end{equation}
Hence, the two successive rotations $R_L^{\dagger}(\hat{{\bf p}}') R_L(\hat{{\bf p}})$ applied to the
 state $|p\hat{{\bf z}} \rangle $ can be replaced by the single rotation $R_L(\hat{{\bf p}}'')$. Consequently any number of successive rotations in momentum space can always be replaced by one rotation with 
the corresponding rotation angles. 
The angles $(\theta '',\phi '')$ of ${\bf p}''$ are determined by the angles $(\theta ,\phi )$ of ${\bf p}$ 
and $(\theta ',\phi ')$ of ${\bf p}'$ by
\begin{eqnarray}
\cos \theta '' & = & \cos \theta \cos \theta '+\sin \theta \sin \theta '\cos (\phi -\phi ') \label{dou6}\\
\sin \theta ''e^{i\phi ''} & = & -\cos \theta \sin \theta '+\sin \theta \cos \theta '\cos (\phi -\phi ') + i \sin \theta \sin (\phi -\phi ') \label{dou7} ,
\end{eqnarray}
and are obtained from the rotation matrices of the Cartesian coordinates, which correspond to the rotations in Eq.~(\ref{dou5}). Such a rotation matrix $M(\alpha \beta \gamma )$ corresponding to $R(\alpha \beta \gamma )$ is given in Eq.~(\ref{dou8}).

\subsection{Two Successive Rotations in Spin Space}

We denote the rotation operator in spin space as $R_S(\hat{{\bf p}})$, which is given in term of the total spin operator ${\bf S}$ as 
\begin{equation}
R_S(\hat{{\bf p}})=R_S(\phi \theta 0)=e^{-iS_{z}\phi }e^{-iS_{y}\theta } .
\end{equation}

The rotation identity given in Eq.~(\ref{dou5}) may not apply in spin space. Therefore, 
we evaluate two successive rotations in spin space, independent of the evaluation in momentum space. 
We compare the rotated spin state or the helicity state $|\hat{{\bf p}}''S\Lambda \rangle
_{\mbox{I}}$ with $|\hat{{\bf p}}''S\Lambda \rangle _{\mbox{II}}$ given by 
\begin{eqnarray}
|\hat{{\bf p}}''S\Lambda \rangle _{\mbox{I}} & = & R_S(\hat{{\bf p}}'') |\hat{{\bf z}}S\Lambda \rangle \label{dou12} \\
|\hat{{\bf p}}''S\Lambda \rangle _{\mbox{II}} & = & R_S^{\dagger}(\hat{{\bf p}}') R_S(\hat{{\bf p}}) |\hat{{\bf z}}S\Lambda \rangle . \label{dou13}
\end{eqnarray}
It should be pointed out that here the relation between $(\theta '',\phi '')$, $(\theta ',\phi ')$ and $(\theta ,\phi )$ given in Eqs.~(\ref{dou6}) and (\ref{dou7}) is still valid, since transformations of the Cartesian coordinates are the same in both momentum space and spin space. 

Both states $|\hat{{\bf p}}''S\Lambda \rangle _{\mbox{I}}$ and $|\hat{{\bf p}}''S\Lambda \rangle _{\mbox{II}}$ are eigenstates of the helicity operator ${\bf S}\cdot \hat{{\bf p}}''$ with eigenvalue $\Lambda $, as can be shown as follows:
\begin{eqnarray}
{\bf S}\cdot \hat{{\bf p}}'' |\hat{{\bf p}}''S\Lambda \rangle _{\mbox{I}} & = & R_S(\hat{{\bf p}}'') {\bf S}\cdot \hat{{\bf z}} R^{\dagger}_S(\hat{{\bf p}}'') R_S(\hat{{\bf p}}'') |\hat{{\bf z}}S\Lambda \rangle  \nonumber\\
& = & R_S(\hat{{\bf p}}'') {\bf S}\cdot \hat{{\bf z}} |\hat{{\bf z}}S\Lambda \rangle \nonumber\\
& = & \Lambda R_S(\hat{{\bf p}}'') |\hat{{\bf z}}S\Lambda \rangle \nonumber\\
& = & \Lambda |\hat{{\bf p}}''S\Lambda \rangle _{\mbox{I}} \\
{\bf S}\cdot \hat{{\bf p}}'' |\hat{{\bf p}}''S\Lambda \rangle _{\mbox{II}} & = & R_S^{\dagger}(\hat{{\bf p}}') R_S(\hat{{\bf p}}) {\bf S}\cdot \hat{{\bf z}} R^{\dagger}_S(\hat{{\bf p}}) R_S(\hat{{\bf p}}') R^{\dagger}_S(\hat{{\bf p}}') R_S(\hat{{\bf p}}) |\hat{{\bf z}}S\Lambda \rangle  \nonumber\\
& = & R_S^{\dagger}(\hat{{\bf p}}') R_S(\hat{{\bf p}}) {\bf S}\cdot \hat{{\bf z}} |\hat{{\bf z}}S\Lambda \rangle  \nonumber\\
& = & \Lambda R_S^{\dagger}(\hat{{\bf p}}') R_S(\hat{{\bf p}}) |\hat{{\bf z}}S\Lambda \rangle  \nonumber\\
& = & \Lambda |\hat{{\bf p}}''S\Lambda \rangle _{\mbox{II}} .
\end{eqnarray}
Moreover, because of the unitarity transformations in Eqs.~(\ref{dou12}) and (\ref{dou13}) the 
two states have the same norm and can at most differ  by a phase factor. 

The helicity states $|\hat{{\bf p}}''S\Lambda \rangle _{\mbox{I}}$ and $|\hat{{\bf p}}''S\Lambda \rangle _{\mbox{II}}$ are expanded in the spin states $|\hat{{\bf z}}S\Lambda \rangle $ as 
\begin{eqnarray}
|\hat{{\bf p}}''S\Lambda \rangle _{\mbox{I}} & = & R_S(\hat{{\bf p}}'') |\hat{{\bf z}}S\Lambda \rangle = \sum _{\Lambda '} |\hat{{\bf z}}S\Lambda '\rangle D^S_{\Lambda '\Lambda }(\phi ''\theta ''0) \\
|\hat{{\bf p}}''S\Lambda \rangle _{\mbox{II}} & = & R_S^{\dagger}(\hat{{\bf p}}') R_S(\hat{{\bf p}}) |\hat{{\bf z}}S\Lambda \rangle \nonumber\\
& = & \sum _{\Lambda 'N} |\hat{{\bf z}}S\Lambda '\rangle \langle\hat{{\bf z}}S\Lambda '| R_S^{\dagger}(\phi '\theta '0) |\hat{{\bf z}}SN\rangle \langle\hat{{\bf z}}SN| R_S(\phi \theta 0) |\hat{{\bf z}}S\Lambda \rangle  \nonumber\\
& = & \sum _{\Lambda '} |\hat{{\bf z}}S\Lambda '\rangle \sum _{N} D^{S\ast}_{N\Lambda '}(\phi '\theta '0) D^S_{N\Lambda }(\phi \theta 0) \nonumber\\
& \equiv & \sum _{\Lambda '} |\hat{{\bf z}}S\Lambda '\rangle X^S_{\Lambda '\Lambda}(\phi ''\theta ''0) ,
\end{eqnarray}
where
\begin{equation}
X^S_{\Lambda '\Lambda}(\phi ''\theta ''0) \equiv \sum _{N} D^{S\ast}_{N\Lambda '}(\phi '\theta '0) D^S_{N\Lambda }(\phi \theta 0) .
\end{equation}
Therefore, instead of comparing $|\hat{{\bf p}}''S\Lambda \rangle _{\mbox{I}} $ with $|\hat{{\bf p}}''S\Lambda \rangle _{\mbox{II}} $ we compare $D^S_{\Lambda '\Lambda}(\phi ''\theta ''0) $ with $X^S_{\Lambda '\Lambda}(\phi ''\theta ''0) $, since these are known functions. We have two spin cases $S = 0$ and $S = 1$. For $S = 0$ the spin state is rotationally invariant and thus we can immediately get  
\begin{equation}\label{dou19}
X^0_{00}(\phi ''\theta ''0) = D^0_{00}(\phi ''\theta ''0) = 1 
\end{equation}
and correspondingly 
\begin{equation}
|\hat{{\bf p}}''00 \rangle _{\mbox{I}} = |\hat{{\bf p}}''00 \rangle _{\mbox{II}} = |\hat{{\bf z}}00 \rangle . 
\end{equation}
For $S = 1$ we make use of a symmetry relation for the Wigner D-functions given as 
\begin{equation}
D^{j\ast}_{m'm}(\alpha \beta \gamma) = (-)^{m'-m}D^j_{-m',-m}(\alpha \beta \gamma) , 
\end{equation}
allowing to leave out the case with initial helicity $\Lambda = -1$ and consider only six cases with $\Lambda ' = 1,0,-1$ and $\Lambda = 1,0$. 

The Wigner D-function $D^1_{\Lambda '\Lambda }(\phi \theta 0) $ is given as 
\begin{equation}\label{dou9}
D^1(\phi \theta 0) = \left(\begin{array}{ccc}
                            e^{-i\phi }\frac{1+\cos \theta }{2} &
                           -e^{-i\phi }\frac{\sin \theta }{\sqrt{2}} &
                            e^{-i\phi }\frac{1-\cos \theta }{2} \\
                            \frac{\sin \theta }{\sqrt{2}} & 
                            \cos \theta & 
                           -\frac{\sin \theta }{\sqrt{2}} \\
                            e^{ i\phi }\frac{1-\cos \theta }{2} &
                            e^{ i\phi }\frac{\sin \theta }{\sqrt{2}} &
                            e^{ i\phi }\frac{1+\cos \theta }{2} 
                           \end{array}\right) .
\end{equation}
For $\Lambda = 0$ it follows that 
\begin{eqnarray}
X^1_{ 10}(\phi ''\theta ''0) = & -e^{-i\phi ''} \frac{\sin \theta ''}{\sqrt{2}} & = D^1_{ 10}(\phi ''\theta ''0) \label{douadd1}\\
X^1_{ 00}(\phi ''\theta ''0) = & \cos \theta '' & = D^1_{ 00}(\phi ''\theta ''0) \\
X^1_{-10}(\phi ''\theta ''0) = & e^{i\phi ''} \frac{\sin \theta ''}{\sqrt{2}} & = D^1_{-10}(\phi ''\theta ''0) \label{douadd2},
\end{eqnarray}
and thus,
\begin{equation}\label{dou10}
X^1_{\Lambda '0}(\phi ''\theta ''0) = D^1_{\Lambda '0}(\phi ''\theta ''0) 
\end{equation}
and correspondingly
\begin{equation}
|\hat{{\bf p}}''10 \rangle _{\mbox{I}} = |\hat{{\bf p}}''10 \rangle _{\mbox{II}} . 
\end{equation}
For $\Lambda = 1$ we obtain 
\begin{eqnarray}
X^1_{ 11}(\phi ''\theta ''0) & = & \frac{1}{2} \left\{ (1+\cos \theta \cos \theta ') \cos (\phi -\phi ') + \sin \theta \sin \theta ' \right\} \nonumber\\
 &  & - \frac{i}{2} (\cos \theta +\cos \theta ') \sin (\phi -\phi ') \label{dou15}\\
X^1_{ 01}(\phi ''\theta ''0) & = & \frac{1}{\sqrt{2}} \left\{ -\cos \theta \sin \theta ' \cos (\phi -\phi ') + \sin \theta \cos \theta ' + i \sin \theta ' \sin (\phi -\phi ') \right\} \\
X^1_{-11}(\phi ''\theta ''0) & = & \frac{1}{2} \left\{ (1-\cos \theta \cos \theta ') \cos (\phi -\phi ') - \sin \theta \sin \theta ' \right\} \nonumber\\
 &  & - \frac{i}{2} (\cos \theta -\cos \theta ') \sin (\phi -\phi ') \label{dou16}.
\end{eqnarray}
Hence, for $\Lambda = 1$ apparently $X^1_{\Lambda '1}(\phi ''\theta ''0)$ differs from $D^1_{\Lambda '1}(\phi ''\theta ''0)$, and correspondingly $|\hat{{\bf p}}''11 \rangle _{\mbox{II}}$ from $|\hat{{\bf p}}''11 \rangle _{\mbox{I}}$, by a phase factor. Now the difference between Eqs.~(\ref{douadd1})-(\ref{douadd2}) and Eqs.~(\ref{dou15})-(\ref{dou16}) is connected to the value of $\Lambda $. Therefore, the phase factor must depend on $\Lambda $ and is independent of $\Lambda '$. The latter can be understood as we see that there is no $\Lambda '$-dependence in Eqs.~(\ref{dou12}) and (\ref{dou13}). In addition the phase factor also depends on the set of angles $(\phi ,\theta ,\phi ',\theta ')$. Thus we write 
\begin{equation}\label{dou17}
X^1_{\Lambda '\Lambda }(\phi ''\theta ''0) = e^{i\Omega \Lambda} D^1_{\Lambda '\Lambda }(\phi ''\theta ''0) ,
\end{equation}
where $\Omega $ depends on the set of angles $(\phi ,\theta ,\phi ',\theta ')$. Noting that
$D^1_{01}(\phi ''\theta ''0)$ is real the phase  $\Omega $ can be given through its tangential as 
\begin{equation}\label{dou22}
\tan \Omega = \frac{Im\{X^1_{01}(\phi ''\theta ''0)\}}{Re\{X^1_{01}(\phi ''\theta ''0)\}} = \frac{\sin \theta ' \sin (\phi -\phi ') }{-\cos \theta \sin \theta ' \cos (\phi -\phi ') + \sin \theta \cos \theta ' }.
\end{equation}
The $\Omega $ calculated in Eq.~(\ref{dou22}) is also valid for other combinations of $\Lambda '$ and $\Lambda $, since $\Omega $ is independent of $\Lambda '$ and $\Lambda $. After all these evaluations we summarize that 
\begin{eqnarray}
R_S^{\dagger}(\hat{{\bf p}}') R_S(\hat{{\bf p}}) |\hat{{\bf z}}S\Lambda \rangle & = & e^{i\Omega \Lambda} R_S(\hat{{\bf p}}'') |\hat{{\bf z}}S\Lambda \rangle \label{dou20}\\
e^{i\Omega \Lambda} & = & \frac{\sum ^{S}_{N=-S}D^{S\ast }_{N\Lambda '}(\phi '\theta '0)D^{S}_{N\Lambda }(\phi \theta 0)}{D^{S}_{\Lambda '\Lambda }(\phi ''\theta ''0)} \label{dou21}. 
\end{eqnarray}
We have restored the spin notation $S$, since Eqs.~(\ref{dou20}) and (\ref{dou21}) are general and hence apply to arbitrary spin $S$, including $S = 0$.



\noindent
\begin{figure}
\caption{\label{ndrfig1} The spin averaged differential cross section  $\frac{d^2 \sigma }{dE_n
d\theta}$ [mb/(MeV $\cdot$ sr)], the analyzing power $A_y$ and the polarization-transfer coefficient
$D_{sl}$ for the (p,n) charge exchange process at projectile energy
 $E_{lab} = 100$~MeV and neutron laboratory scattering angle $\theta _{lab} = 13^o$. 
The solid line represents the 3D calculation, the dashed line the PW calculation with
$j = 7,\, J = 31/2$. Both calculations are based on the Bonn-B potential~\protect\cite{16}. }
\end{figure}

\noindent
\begin{figure}
\caption{\label{ndrfig2}Same as Fig.~\ref{ndrfig1}, but for projectile energy 
$E_{lab} = 197$~MeV. The solid line
represents the 3D calculation, the dashed and dotted lines represent PW calculations with different
2N total angular momentum $j$ and total 3N angular momentum $J$ as indicated in the figure. All
calculations are based on the Bonn-B potential. }
\end{figure}

\noindent
\begin{figure}
\caption{\label{ndrfig4}
The spin averaged differential cross section  $\frac{d^2 \sigma }{dE_n
d\theta}$ [mb/(MeV $\cdot$ sr)], the analyzing power $A_y$ and the polarization-transfer coefficient
$D_{ll}$ for the (p,n) charge exchange process at projectile energy
 $E_{lab} = 197$~MeV and neutron laboratory scattering angle $\theta _{lab} = 24^o$.
The solid (short dashed) line represent the 3D calculations for the first order term based on the
AV18 (Bonn-B) potential. The long-dashed (dotted) lines stand for the partial wave based, full
Faddeev calculations based on the AV18 (Bonn-B) potentials. The data are taken from Ref.~\protect\cite{16}.
}
\end{figure}

\noindent
\begin{figure}
\caption{\label{ndrfig5}The same as in Fig.~\ref{ndrfig4}, but for the spin averaged differential cross section $\frac{d^2 \sigma }{dE_n d\theta}$ [mb/(MeV.sr)], 
the analyzing power $A_y$ and the polarization-transfer coefficient $D_{ss}$ at $\theta _{lab} = 37^o$.}
\end{figure}

\noindent
\begin{figure}
\caption{\label{ndrfig7} 
The spin averaged differential cross section  $\frac{d^2 \sigma }{dE_n   
d\theta}$ [mb/(MeV $\cdot$ sr)] and the analyzing power $A_y$ for the (p,n) charge exchange process at
projectile energy $E_{lab} = 100$~MeV and neutron laboratory scattering angle $\theta _{lab} = 24^o$
The solid (short dashed) line give  the nonrelativistic  3D calculations for the first order term
based on the AV18 (Bonn-B) potential. The long-dashed (dotted) lines represent the 3D calculations that
include the effects of relativistic kinematics. 
}
\end{figure}

\noindent
\begin{figure}
\caption{\label{ndrfig8} The spin averaged differential cross section  $\frac{d^2 \sigma }{dE_n
d\theta}$ [mb/(MeV $\cdot$ sr)], the analyzing power $A_y$ and the polarization-transfer coefficient
$D_{ll}$ for the (p,n) charge exchange process at projectile energy
 $E_{lab} = 197$~MeV and neutron laboratory scattering angle $\theta _{lab} = 24^o$.
The solid (short dashed) line give  the nonrelativistic  3D calculations for the first order term   
based on the AV18 (Bonn-B) potential. The long-dashed (dotted) lines represent the 3D calculations that 
include the effects of relativistic kinematics. The data are taken from Ref.~\protect\cite{16}.
}
\end{figure}

\noindent
\begin{figure}
\caption{\label{ndrfig9}The same as in Fig.~\ref{ndrfig8},  but for the spin averaged differential cross
section $\frac{d^2 \sigma }{dE_n d\theta}$ [mb/(MeV.sr)], the analyzing power $A_y$ and the polarization-transfer coefficient $D_{ss}$ at $\theta _{lab} = 37^o$.}
\end{figure}

\noindent
\begin{figure}
\caption{\label{ndrfig11} The spin averaged differential cross section  $\frac{d^2 \sigma }{dE_n
d\theta}$ [mb/(MeV $\cdot$ sr)], the analyzing power $A_y$ and the polarization-transfer coefficient
$D_{sl}$ for the (p,n) charge exchange process at projectile energy
 $E_{lab} = 346$~MeV and neutron laboratory scattering angle $\theta _{lab} = 22^o$.
The solid (short dashed) line give  the nonrelativistic  3D calculations for the first order term
based on the AV18 (Bonn-B) potential. The long-dashed (dotted) lines represent the 3D calculations that
include the effects of relativistic kinematics. The data are taken from Ref.~\protect\cite{ndx346}.
}
\end{figure}

\noindent
\begin{figure}
\caption{\label{ndrfig12} The spin averaged differential cross section  $\frac{d^2 \sigma }{dE_n
d\theta}$ [mb/(MeV $\cdot$ sr)], the analyzing power $A_y$ and the polarization-transfer coefficient
$D_{ll}$ for the (p,n) charge exchange process at projectile energy
 $E_{lab} = 495$~MeV and neutron laboratory scattering angle $\theta _{lab} = 18^o$.
The solid (short dashed) line give  the nonrelativistic  3D calculations for the first order term
based on the AV18 (Bonn-B) potential. The long-dashed (dotted) lines represent the 3D calculations that
include the effects of relativistic kinematics. The data are taken from Ref.~\protect\cite{ndx495}.
}
\end{figure}


\newpage
\centerline{Fig. 1}\vspace*{-3mm}
\begin{figure}
{\par 
\begin{center}
\input{Xd213xpwbb.100.tex}\vspace*{-3mm}
\end{center}
\par}
{\par 
\begin{center}
\input{Xay13xpwbb.100.tex}\vspace*{-3mm}
\end{center}
\par}
{\par 
\begin{center}
\input{Xdsl13xpwbb.100.tex}
\end{center}
\par}
\end{figure}

\newpage
\centerline{Fig. 2}\vspace*{-3mm}
\begin{figure}
{\par 
\begin{center}
\input{Xd213xpwbb.197.tex}\vspace*{-3mm}
\end{center}
\par}
{\par 
\begin{center}
\input{Xay13xpwbb.197.tex}\vspace*{-3mm}
\end{center}
\par}
{\par 
\begin{center}
\input{Xdsl13xpwbb.197.tex}
\end{center}
\par}
\end{figure}

\newpage
\centerline{Fig. 3}\vspace*{-3mm}
\begin{figure}
{\par 
\begin{center}
\input{Xd224xfull.197.tex}\vspace*{-3mm}
\end{center}
\par}
{\par 
\begin{center}
\input{Xay24xfull.197.tex}\vspace*{-3mm}
\end{center}
\par}
{\par 
\begin{center}
\input{Xdll24xfull.197.tex}
\end{center}
\par}
\end{figure}

\newpage
\centerline{Fig. 4}\vspace*{-3mm}
\begin{figure}
{\par 
\begin{center}
\input{Xd237xfull.197.tex}\vspace*{-3mm}
\end{center}
\par}
{\par 
\begin{center}
\input{Xay37xfull.197.tex}\vspace*{-3mm}
\end{center}
\par}
{\par 
\begin{center}
\input{Xdss37xfull.197.tex}
\end{center}
\par}
\end{figure}

\newpage
\centerline{Fig. 5}
\begin{figure}
{\par 
\begin{center}
\input{Xd224xrel.100.tex}\vspace*{-5mm}
\end{center}
\par}
{\par 
\begin{center}
\input{Xay24xrel.100.tex}
\end{center}
\par}
\end{figure}

\newpage
\centerline{Fig. 6}\vspace*{-3mm}
\begin{figure}
{\par 
\begin{center}
\input{Xd224xrel.197.tex}\vspace*{-3mm}
\end{center}
\par}
{\par 
\begin{center}
\input{Xay24xrel.197.tex}\vspace*{-3mm}
\end{center}
\par}
{\par 
\begin{center}
\input{Xdll24xrel.197.tex}
\end{center}
\par}
\end{figure}

\newpage
\centerline{Fig. 7}\vspace*{-3mm}
\begin{figure}
{\par 
\begin{center}
\input{Xd237xrel.197.tex}\vspace*{-3mm}
\end{center}
\par}
{\par 
\begin{center}
\input{Xay37xrel.197.tex}\vspace*{-3mm}
\end{center}
\par}
{\par 
\begin{center}
\input{Xdss37xrel.197.tex}
\end{center}
\par}
\end{figure}

\newpage
\centerline{Fig. 8}\vspace*{-3mm}
\begin{figure}
{\par 
\begin{center}
\input{Xd222xrel.346.tex}\vspace*{-3mm}
\end{center}
\par}
{\par 
\begin{center}
\input{Xay22xrel.346.tex}\vspace*{-3mm}
\end{center}
\par}
{\par 
\begin{center}
\input{Xdsl22xrel.346.tex}
\end{center}
\par}
\end{figure}

\newpage
\centerline{Fig. 9}\vspace*{-3mm}
\begin{figure}
{\par 
\begin{center}
\input{Xd218xrel.495.tex}\vspace*{-3mm}
\end{center}
\par}
{\par 
\begin{center}
\input{Xay18xrel.495.tex}\vspace*{-3mm}
\end{center}
\par}
{\par 
\begin{center}
\input{Xdll18xrel.495.tex}
\end{center}
\par}
\end{figure}

\end{document}